\documentclass[pre,aps,amsmath,amssymb,superscriptaddress]{revtex4-2}

\usepackage{graphicx} 
\usepackage{color}
\usepackage{bm}
\usepackage{blindtext}
\usepackage{enumerate}
\usepackage{enumitem}
\usepackage{scalerel} 
\usepackage{subfig}
\usepackage{appendix}

\def\tU{{\scaleto{U}{3pt}}}
\def\tF{{\scaleto{F}{3pt}}}
\def\tC{{\scaleto{C}{3pt}}}
\def\tT{{\scaleto{T}{3pt}}}
\def\x{{\bm x}}

\def\u{{\bm u}}
\def\bn{{\bm n}}

\def\W{{\bm W}}

\def\bxi{{\bm{\xi}}}

\def\F{{{\bm F}}}

\def\d{{{\mbox{d}}}}

\newtheorem{remark}{Remark}

\begin{document}
\title{A unified gas-kinetic framework from Boltzmann to Navier-Stokes scales}

\author{Zhaoli Guo}
\email[Corresponding Author: ]{zlguo@hust.edu.cn}
\affiliation{Institute of Multidisciplinary Research for Mathematics and Applied Science, Huazhong University of Science and Technology, Wuhan 430074, China}
\affiliation{State Key Laboratory of Coal Combustion, Huazhong University of Science and Technology, Wuhan 430074, China}

\author{Kun Xu}
\email{makxu@ust.hk}
\affiliation{Department of Mathematics, Hong Kong University of Science and Technology, Clear Water Bay, Hong Kong, China}

\author{Yajun Zhu}
\email{yajun\_zhu@163.com}
\affiliation{Research and Development Office, Shanghai Suochen Information Technology Co., Ltd, Shanghai 200021, China}

\begin{abstract}
Neither molecular kinetics nor continuum fluid dynamics alone is adequate to describe multiscale gas flows across different regimes. Bridging these regimes within a single self-consistent framework has long been a central challenge in fluid mechanics. We propose a unified gas kinetic framework that classifies molecules by their collision histories over an observation timescale. This formulation recovers the Boltzmann and Navier–Stokes equations as limiting cases, providing a transparent connection between kinetic and hydrodynamic descriptions. Beyond practical advantages for multiscale modeling, this framework offers a new perspective on Hilbert’s sixth problem by linking microscopic dynamics to continuum mechanics through a tunable observational scale.
\end{abstract}


\maketitle

\section{Introduction}
From the upper atmosphere encountered by re-entering spacecraft to the micro channels of microfluidic devices, gases frequently exhibit behavior that spans multiple physical regimes. The absence of distinct time and/or length scale separations, caused by the strong coupling of flow physics across different scales, makes it challenging to describe multiscale fluid dynamics within a unified framework. At one extreme, the Navier–Stokes equations provide an efficient and accurate description when molecular collisions are frequent and the gas behaves as a continuum \cite{ref:LandauBook}. 
At the other limit, the Boltzmann equation \cite{ref:CercignaniBook} governs kinetic-scale dynamics characterized by the molecular mean free path and mean collision time. It is formally valid from the collisionless free-molecular limit through the transitional regime to the continuum limit, and can in principle yield accurate solutions by fully resolving the kinetic-scale dynamics everywhere. However, the direct use of the Boltzmann equation in the transitional regime and the continuum regime is often computationally prohibitive by imposing the kinetic scale resolution. Conversely, the Navier–Stokes equations lose accuracy in the transitional regime, where the mean free path is comparable to the characteristic flow scale. Capturing such flows in a single, predictive, and efficient framework therefore remains a central challenge in nonequilibrium gas dynamics and a core difficulty in multiscale modeling.
This difficulty is closely tied to Hilbert’s sixth problem, posed in 1900 \cite{ref:Hilbert1901}, which seeks a rigorous connection between atomistic models and the continuum laws of mechanics. Over a century later, the lack of a fully general and physically consistent bridge between these levels continues to limit both theoretical understanding and practical simulation capabilities \cite{ref:Gorban2018}.

In the Boltzmann equation, all molecules are assumed to undergo binary collisions, and the nonequilibrium dynamics is described by the evolution of a single velocity distribution function. Nevertheless, within a finite observation time, not all molecules actually experience collisions. Thus, describing the system solely with a single distribution function cannot distinguish molecules with different collision histories. In essence, while the Boltzmann equation is exact at the kinetic scale, it cannot directly resolve the distinct dynamical behaviors of molecules that do or do not collide within the chosen observation window.
Moreover, physical laws governing gas transport must be formulated with respect to a specific observation scale. At the kinetic scale, molecular motions and collisions dominate; at the hydrodynamic scale, the dynamics are governed by collective fluid properties, and the details of individual molecular trajectories are irrelevant; at a numerical scale, the observed dynamics depend on the ratio between the grid size and mean free path \cite{ref:UP2023}. Introducing an explicit observation time (and associated length) scale is therefore essential to properly describe multiscale transport processes.

Motivated by this perspective, we propose a unified gas kinetic framework (UGKF) for multiscale flows based on the classification of molecules by their collision histories. This idea is inspired by our recent analysis of the unified gas-kinetic wave-particle method \cite{ref:WP_KineticRep,ref:WP_LiuJCP2020}, where collisional and free-transport components were distinguished to interpret the method’s multiscale behavior. Here we extend the idea to construct a general set of scale-dependent kinetic equations, which can be regarded as a generalized gas dynamic model for multiscale flow physics.
Specifically, we embed an observation scale $h$ between the kinetic and hydrodynamic limits. Within this window, molecules are naturally divided into three populations: (i) free-transport molecules, which remain collisionless throughout $(0,h]$; (ii) transitional molecules, which collide after a free-flight of duration $t<h$; and (iii) collided molecules, which have already experienced collisions within $(0,t]$. Then scale-dependent kinetic equations are derived for each of these populations from the Boltzmann equation, providing a framework that continuously transitions between the kinetic and hydrodynamic limits as $h$ varies. 
This UGKF not only provides deeper insight into the transitional physics with the variation of scales, but also reveals the internal structure of the Boltzmann equation. Particularly, the explicit scale-dependent decomposition offers a possible novel route toward addressing Hilbert’s sixth problem by constructing a consistent bridge between microscopic collision dynamics and macroscopic fluid behavior.

The rest of this paper is organized as follows. In Sec.2, the unified gas kinetic framework is derived from the Boltzmann equation, and the multiscale nature of the system is analyzed in Sec. 3, then Sec. 4 provides several benchmark tests and applications to illustrate the effectiveness of the UGKF, and finally a brief summary is presented in Sec. 5.

\section{Unified gas kinetic framework}
\label{sec:BE}
Our starting point is the Boltzmann equation for monatomic gases \cite{ref:ChapmanBook},
\begin{equation}
	\label{eq:BE}
	\partial_t f +\bxi\cdot \nabla f = Q(f,f_*),
\end{equation}
where $f(\x,\bxi,t)$ is the velocity distribution function for gas molecules moving with velocity $\bxi$ at position $\x$ and time $t$, and $Q$ is the collision operator defined as a quadratic integral modeling the binary elastic collision between molecules
\begin{equation}
	\label{eq:Q}
	Q(f,f_*)=\int_{\mathbb{R}^3}\int_{\mathbb{S}^2}{B(|\bm{v}_r|,\cos\theta)[f(\bxi_*')f(\bxi')-f(\bxi_*)f(\bxi)]d\bm{\omega} d\bxi_*},
\end{equation}
where $(\bxi,\bxi_*)$ and $(\bxi',\bxi_*')$ are the velocity pairs before and after the collision, during which the momentum and energy are
conserved, with $\bxi'=\bxi-(\bm{\omega}\cdot\bm{v}_r)\bm{\omega}$ and $\bxi_*'=\bxi_*+(\bm{\omega}\cdot\bm{v}_r)\bm{\omega}$. Here, $\bm{\omega}$ is the scattering direction varying on the unit sphere, $B$ is the non-negative collision kernel function, $\bm{v}_r=\bxi-\bxi_*$ is the relative velocity, and $\cos\theta=\bm{\omega}\cdot\bm{v}_r/|\bm{v}_r|$. 
It should be noted that the Boltzmann equation is inherently defined at the kinetic scale, characterized by the mean collision time $\tau$ and mean free path $\lambda$, and all molecules are assumed to undergo collisions. 


The conservative hydrodynamic variables $\W=(\rho, \rho\u, \rho E)$ are defined as the conservative moments of the distribution function $f$,
\begin{equation}
	\W=\int{\bm{\psi(\bxi)}f d\bxi}, \quad \bm{\psi}=(1,\bxi, \tfrac{1}{2}\xi^2).
\end{equation}
Here $\rho$, $\u$, and $E=\tfrac{1}{2}u^2+\tfrac{3}{2}RT$ are the density, velocity, and total energy of the gas, respectively, with $R$ the gas constant and $T$ the temperature. 

If the total cross section of the collision is finite, the collision operator can be split into a gain term $Q^+$ and a loss term $\nu f$ \cite{ref:CercignaniBook,ref:Villani2002review}, i.e., 
\begin{equation}
	Q(f,f_*)=Q^+(f,f_*)-\nu f,
\end{equation}
where 
\begin{equation}
	Q^+(\bxi)=\int_{\mathbb{R}^3}\int_{\mathbb{S}^2}{B(|\bm{v}_r|,\cos\theta)f(\bxi_*')f(\bxi')d\bm{\omega} d\bxi_*},
\end{equation}
and $\nu(\bxi)$ is the collision frequency,
\begin{equation}
	\nu(\bxi)=\int_{\mathbb{R}^3}\int_{\mathbb{S}^2}{B(|\bm{v}_r|,\cos\theta)f(\bxi_*)d\bm{\omega} d\bxi_*}.   
\end{equation}
It is clear that the gain term $Q^+$ represents the distribution function for the post-collision molecules. 
In this case, the Boltzmann equation \eqref{eq:BE} admits a formal analytical solution,
\begin{equation}
	\label{eq:BE-solution}
	f(\x,\bxi,t)=\underbrace{e^{-\bar{\nu}(\x,\bxi,t)} f(\x_0,\bxi,0)}_{f_\tU(\x,\bxi,t)} + \underbrace{\int_0^t {\left[e^{-(\bar{\nu}(\x,\bxi,t)-\bar{\nu}(\x',\bxi,t'))} Q^+(\x',\bxi, t') \right]\d t'}}_{f_\tC(\x,\bxi,t)},
\end{equation}
where $\x'=\x_0+\bxi t'$ and $\bar{\nu}(\x,\bxi,t)=\int_0^t{[\nu\left(\x_0 +\bxi s,\bxi,s\right)]\d s}$, with $\x_0=\x-\bxi t$ being the starting point of the characteristic line. Note that $\bar{\nu}(\x,\bxi,t)$ represents the number of collisions of molecules moving with velocity $\bxi$ along the characteristic line from time $0$ to $t$. 

The formal solution \eqref{eq:BE-solution} suggests that the distribution $f$ can be decomposed into two parts, $f=f_\tU+f_\tC$, where $f_\tU(\x,\bxi,t)$ represents the distribution function of {\em uncollided} molecules transported from $0$ to $t$, while $f_\tC(\x,\bxi,t)$ represents that for {\em collided} molecules already experienced collisions within this time interval. It is noted that the idea of uncollided-collided decomposition of the distribution function was first proposed for the linear neutron transport equation in Ref. \cite{ref:Coll-Hybrid1971} to develop numerical methods with reduced ray effects, and was recently extended to solve the Boltzmann-BGK equation \cite{ref:Coll-Hybrid-BGK2024}. However, in these studies, collision decomposition is proposed intuitively and the aim is to design numerical methods rather than to reveal the underlying physics. 

From the definition of $f_\tU$ given in Eq. \eqref{eq:BE-solution}, it follows that 
$f_\tU(\x,\bxi,t) \to 0$ as $t\to\infty$, suggesting that all uncollided molecules eventually undergo collisions. For a finite observation time $h$, however, there must still be some uncollided molecules present at time $h$. Specifically, a fraction $e^{-\bar{\nu}(\x,\bxi,h)}$ of the initial molecules remain collision-free over $(0,h]$. Consequently, $f_\tU$ can be further decomposed into two parts, 
\begin{equation}
	\label{eq:fU-decompose}
	f_\tU(\x,\bxi,t)=\underbrace{e^{-\bar{\nu}(\x_0+\bxi h,\bxi,h)} f(\x_0,\bxi,0)}_{f_\tF(\x,\bxi,t;h)} +\underbrace{\left(e^{-\bar{\nu}(\x,\bxi,t)}-e^{-\bar{\nu}(\x_0+\bxi h,\bxi,h)}\right) f(\x_0,\bxi,0)}_{f_\tT(\x,\bxi,t;h)},
\end{equation}
where $f_\tF$ represents the {\em free transport} molecules that undergo no collisions during the entire interval $(0,h]$, whereas $f_\tT$ denotes the {\em transitional} molecules that move freely up to time $t$ but collide before $h$. With this observation time scale, the formal solution \eqref{eq:BE-solution} can be rewritten as
\begin{equation}
	\label{eq:BE-Structure}
	f(\x,\bxi,t)=f_\tF(\x,\bxi,t;h)+f_\tT(\x,\bxi,t;h)+f_\tC(\x,\bxi,t).
\end{equation}
Accordingly, the hydrodynamic variables can also be decomposed as $\W=\W_\tF+\W_\tT+\W_\tC$.

\begin{figure}
	\centering
	\subfloat[]{\includegraphics[width=0.4\textwidth]{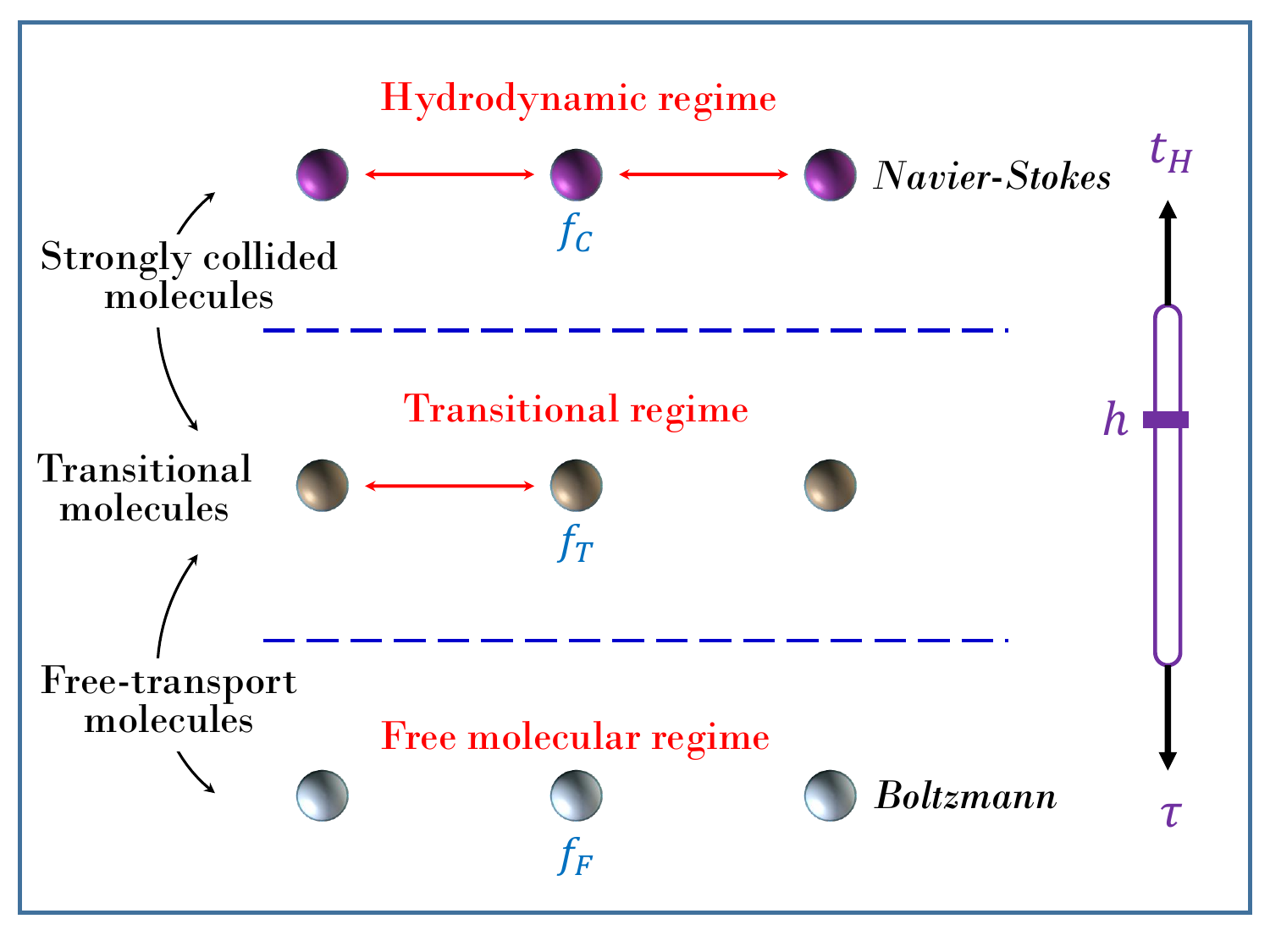}}    
	\subfloat[]{\includegraphics[width=0.4\textwidth]{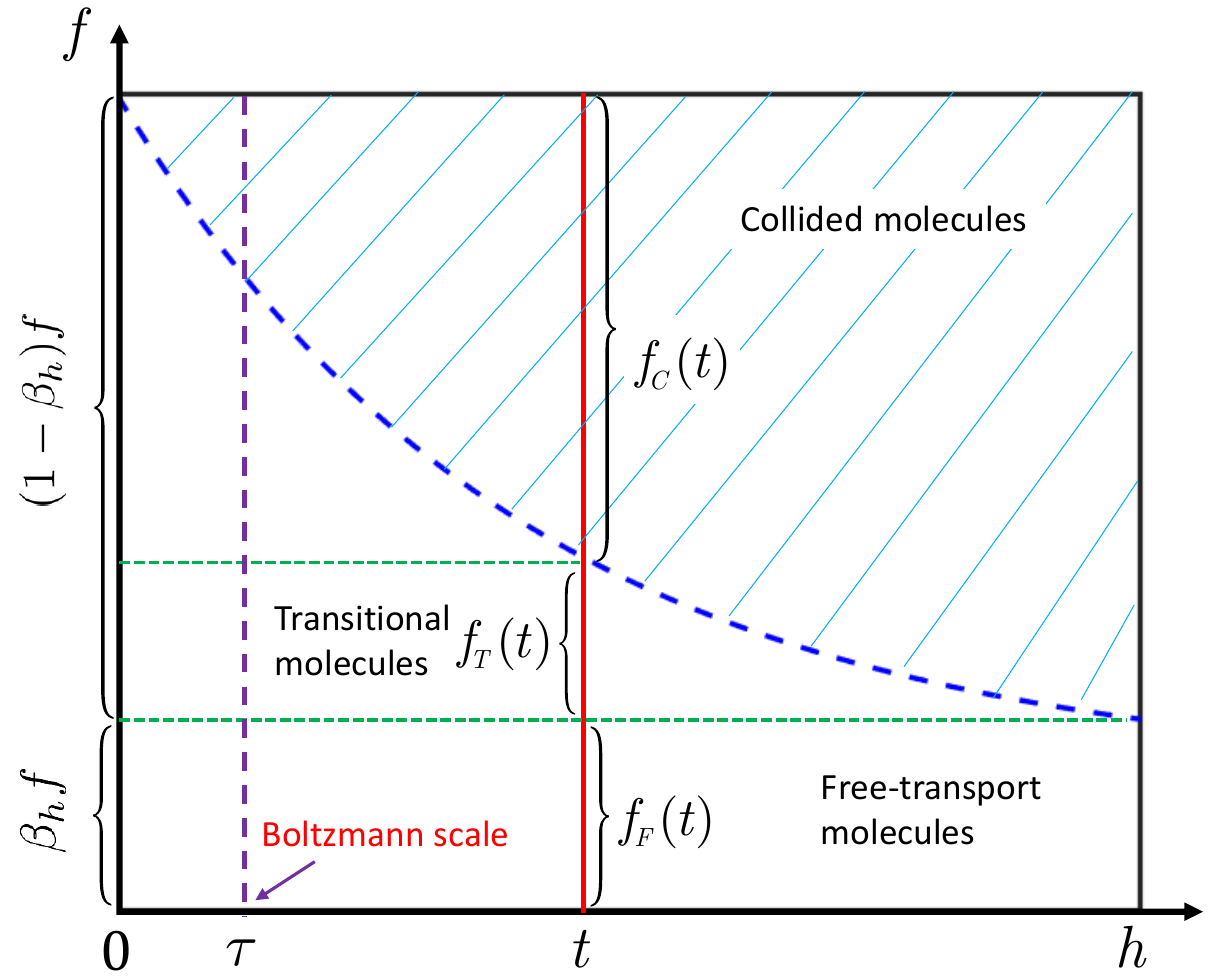}}    
	\caption{(a) Classification of gas molecules based on collisional history and the dominant mechanicsm with the observation time scale $h$ ; (b) Time evolutions of the distribution functions of the three populations. Blue dashed line: The first-collision time of molecules; Shading region: collided molecules; Blank region: Uncollided molecules. $\tau$ is the kinetic time scale characterizing the mean collision time, and $t_H$ is the hydrodynamic time scale characteristizing macroscopic flow evolution over the characteristic length $L_0$.}
	\label{fig:timeFUC} 
\end{figure}

Equation \eqref{eq:BE-Structure} reveals the structure of the solution with the observation time scale $h$, which suggests that the distribution function naturally decomposes into three parts according to molecular collision histories. On this basis and starting from the Boltzmann equation \eqref{eq:BE}, we can obtain the following collision-decomposed kinetic system with well-defined initial conditions, 
\begin{subequations}
	\label{eq:3Population}
	\begin{equation}
		\label{eq:fF}
		\left\{
		\begin{aligned}
			&\partial_t f_{\tF} +\bxi\cdot \nabla f_{\tF} = 0, \quad 0 < t \le h, \\
			& f_{\tF}(\x,\bxi,0;h) = \beta_h(\x,\bxi) f_0(\x,\bxi), \\
		\end{aligned}
		\right.
	\end{equation}
	\begin{equation}
		\label{eq:fT}
		\left\{
		\begin{aligned}
			&\partial_t f_{\tT} +\bxi\cdot \nabla f_{\tT} = -\nu (f_{\tT}+f_{\tF}), \quad 0 < t \le h, \\
			& f_{\tT}(\x,\bxi,0;h) =\left[1-\beta_h(\x,\bxi)\right]f_0(\x,\bxi),\\
		\end{aligned}
		\right.
	\end{equation}
	\begin{equation}
		\label{eq:fC}
		\left\{
		\begin{aligned}
			&\partial_t f_{\tC} +\bxi\cdot \nabla f_{\tC} = -\nu f_{\tC}+Q^+, \quad 0 < t \le h, \\
			& f_{\tC}(\x,\bxi,0) = 0,\\
		\end{aligned}
		\right.
	\end{equation}
\end{subequations}
where $\beta_h(\x,\bxi)=e^{-\bar{\nu}(\x+\bxi h,\bxi,h)}$. The system \eqref{eq:3Population} clearly describes the transport processes of the three types of molecules, which are sketched in Fig. \ref{fig:timeFUC}. Particularly, the initial conditions for the sub-kinetic equations in the system indicate that all molecules are initially uncollided. Among them, a fraction $\beta_h$ undergoes free transport throughout  $0<t\le h$, while the remaining fraction $1-\beta_h$ experiences collisions and transitions to the collided state, as represented by the source terms in Eqs. \eqref{eq:fT} and \eqref{eq:fC}. In addition, the relaxation terms in Eqs. \eqref{eq:fF} and \eqref{eq:fT} reveal that the probability of a molecule avoiding collisions over $(0,t]$ decays as $e^{-\bar{\nu}(t)}$, implying that the distribution of first-collision times follows approximately an exponential law. 

The system \eqref{eq:3Population} constitutes our unified gas kinetic framework (UGKF) governing the multiscale transport ranging from kinetic scale (mean collision time $\tau$ and mean-free-path $\lambda$) to hydrodynamic scale (characteristic time $t_H$ and length $L$), depending on the observation time scale $h$. As will be shown later, the system is dominated by Eq. \eqref{eq:fF} as $h\ll \tau$ (free-molecular regime), suggesting that only free transport process is observed on this scale; On the other hand, as $h\gg \tau$ (continuum regime), intensive collisions happen and the system is dominated by Eq. \eqref{eq:fC}, which can be further approximated by the Navier-Stokes equations; and in between as $h\sim\tau$ (transitional regime),  the three equations are all important and the whole system is equivalent to the Boltzmann equation.

It is straightforward to verify that the formal analytical solutions of the UGKF  \eqref{eq:3Population} coincide with those given by Eqs. \eqref{eq:BE-solution} and \eqref{eq:fU-decompose}. This means that the model \eqref{eq:3Population} is a generalization of the Boltzmann equation \eqref{eq:BE}. Specifically, if $f_\tF$, $f_\tT$, and $f_\tC$ satisfy Eqs. \eqref{eq:fF}, \eqref{eq:fT}, and \eqref{eq:fC}, respectively, their sum $f=f_\tF+f_\tT+f_\tC$ is a solution of the Boltzmann equation \eqref{eq:BE}; Conversely, a solution $f$ of the Boltzmann equation \eqref{eq:BE} can be decomposed into components that satisfy the sub-equations of \eqref{eq:3Population}.
Nonetheless, the distinction between the UGKF \eqref{eq:3Population} and the Boltzmann equation \eqref{eq:BE} is clear and fundamental. The Boltzmann equation describes the evolution of a single distribution function for {\em all} molecules, irrespective of their collision history. In contrast, the present model partitions molecules into three populations according to their collisional behavior within the observation time scale $h$, and tracks their evolution individually. This decomposition provides not only physical clarity but also intrinsic adaptability: by adjusting the modeling scale $h$, the framework naturally resolves transport phenomena across different temporal regimes, as will be demonstrated in next section.

\begin{remark}
	The collision frequency $\nu$ depends on molecular velocity $\bxi$ in general. For Maxwell molecules $\nu$ is independent of $\bxi$ and the kinetic system \eqref{eq:3Population} can be viewed as a decomposition of the Boltzmann-BGK equation. 
\end{remark}

\section{Multiscale nature of the UGKF}
\label{sec:Asymptotic}
\subsection{Asymptotic analysis}
The multiscale properties of the UGKF \eqref{eq:3Population} can be investigated by analyzing the asymptotic limits of the sub-kinetic equations with varying observation time scale.  For this purpose, we first recall the formal analytical solutions,
\begin{subequations}
	\label{eq:AnaSolution}
	\begin{equation}
		\label{eq:AfF}
		f_\tF(\x,\bxi,t;h)=\beta_h(\x_0,\bxi) f(\x_0,\bxi,0), 
	\end{equation}
	\begin{equation}
		\label{eq:AfT}
		f_\tT(\x,\bxi,t) = \left[e^{-\bar{\nu}(\x,\bxi,t))} - \beta_h(\x_0,\bxi)\right] f(\x_0,\bxi,0),
	\end{equation}
	\begin{equation}
		\label{eq:AfC}
		f_\tC(\x,\bxi,t)= \int_0^t {\left[e^{-(\bar{\nu}(\x,\bxi,t)-\bar{\nu}(\x',\bxi,t'))} Q^+(\x',\bxi, t') \right]\d t'}, \quad \x'=\x_0 +\bxi t'.
	\end{equation}
\end{subequations}
Note that 
\begin{equation}
	\begin{aligned}
		\bar{\nu}(\x_0+\bxi h,\bxi,h) &=\int_0^h{[\nu\left(\x_0 +\bxi s,\bxi,s\right)]\d s}\approx h\nu_m , \\    
		\bar{\nu}(\x,\bxi,t) &=\int_0^t{[\nu\left(\x_0+\bxi s),\bxi,s\right)]\d s}\approx t\nu_m,\\ 
		\bar{\nu}(\x,\bxi,t)-\bar{\nu}(\x',\bxi,t') &=\int_{t'}^t{[\nu\left(\x_0+\bxi s,\bxi,s\right)]\d s} \approx (t-t')\nu_m,
	\end{aligned}
\end{equation}
where $\nu_m$ is the mean collision frequency along the characteristic line, and we have assumed that $\nu$ does not change much along the characteristic line over the whole time interval $(0,h]$. With the above estimation, we can obtain that 
\begin{subequations}
	\label{eq:f-beta}
	\begin{equation}
		f_\tF(\x,\bxi,t;h) \approx e^{-h\nu_m} f(\x_0,\bxi,0),\\
	\end{equation}
	\begin{equation}
		f_\tT(\x,\bxi,t;h) \approx \left(e^{-t\nu_m}-e^{-h\nu_m}\right) f(\x_0,\bxi,0) , \\ 
	\end{equation}
	\begin{equation}
		f_\tC(\x,\bxi,t) \approx \left(1-e^{-\nu_m t}\right) f^+(\x,\bxi,t) - \left(\dfrac{1}{\nu_m}(1-e^{-\nu_m t})-te^{-\nu_m t}\right) D_t f^+(\x,\bxi,t),
	\end{equation}
\end{subequations}
for $0<t\le h$, where $f^+=Q^+/\nu_m$ and $D_t=\partial_t+\bxi\cdot\nabla$, and we have assumed in Eq. \eqref{eq:AfC}  that
\begin{equation}
	Q^+(\x',\bxi,t')\approx Q^+(\x,\bxi,t) -(t-t') D_t Q^+(\x,\bxi,t)
\end{equation}

Equation \eqref{eq:f-beta} suggests that the three distribution functions depend on the parameter $\beta_h\approx e^{-\nu_m h}$. Accordingly, the asymptotic behavior of the kinetic system \eqref{eq:3Population} can be analyzed in terms of the parameter $\delta_h=h\nu_m$, which measures the average number of collisions over $(0,h]$. Equivalently, $\delta_h$ is  proportional to the inverse of local Knudsen number determined by the mean collision time $\tau_m=\nu_m^{-1}$ and the observation time scale $h$, i.e., $\delta_h \sim h/\tau_m$. 

We first consider the case $\delta_h\gg 1$, where the observation time scale $h$ is much larger than the mean collision time $\tau_m$. In this regime, collisions are frequent, and the flow observed at this scale is continuum. Specifically, we have
\begin{equation}
	e^{-\nu_m h}=e^{-\delta_h} \to 0, \quad e^{-\nu_m t}=e^{-\delta_h (t/h)} \to 0.
\end{equation}
Therefore, from Eq. \eqref{eq:f-beta} we have
\begin{subequations}
	\begin{equation}
		f_\tF(\x,\bxi,t;h) \to 0, \quad f_\tT(\x,\bxi,t;h) \to 0,
	\end{equation}
	\begin{equation}
		\label{eq:Fc-continuum}
		f_\tC(\x,\bxi,t) \to f^+(\x,\bxi,t)-\dfrac{1}{\nu_m} D_t f^+(\x,\bxi,t).
	\end{equation}   
\end{subequations}
This clearly show that, in the continuum limit, the uncollided molecules, consisting of free-transport and transitional molecules, are negligible during the time interval $(0, h]$.  Furthermore, the system is remains close to equilibrium, and thus 
\begin{equation}
	\label{eq:Q-fM}
	\begin{aligned}
		Q^+(\bxi) & \approx\int_{\mathbb{R}^3}\int_{\mathbb{S}^2} {B(|\bm{v}_r|,\cos\theta)f_M(\bxi_*')f_M(\bxi')d\bm{\omega} d\bxi_*}\\
		& =\int_{\mathbb{R}^3}\int_{\mathbb{S}^2} {B(|\bm{v}_r|,\cos\theta)f_M(\bxi_*)f_M(\bxi)d\bm{\omega} d\bxi_*}\\
		&=\nu_M f_M(\bxi),
	\end{aligned}   
\end{equation}
where $\nu_M$ is the collision frequency at equilibrium, and $f_M$ is the local Maxwellian,
\begin{equation}
	f_M(\x,\bxi,t)=\dfrac{\rho}{(2\pi RT)^{3/2}}\exp\left(-\dfrac{C^2}{2RT}\right), 
\end{equation}
with $\bm{C}=\bxi-\u$ being the peculiar velocity.
Near equilibrium, we can assume $\nu_M(\bxi)\approx \nu_m (\bxi)$, and then $f^+=Q^+/\nu_m\approx f_M$. With this approximation, we can obtain from Eq. \eqref{eq:Fc-continuum} that
\begin{equation}
	\label{eq:Fc-Maxwell}
	f_\tC(\x,\bxi,t) \approx f_M(\x,\bxi,t)-\dfrac{1}{\nu_m} D_t f_M(\x,\bxi,t),
\end{equation}   
which is the Chapman-Enskog approximation of the distribution of $f(\x,\bxi,t)$ at the Navier-Stokes level. 
These arguments suggest that, in the continuum limit, the sub-kinetic equation \eqref{eq:fC} for collided molecules dominates the entire kinetic system \eqref{eq:3Population}, so that  $f\approx f_{\tC}$, which reduces to the Chapman-Enskog solution at the Navier-Stokes level. Note that in the derivation of the macroscopic governing equations in the continuum flow regime, we do not need to explicitly use the Chapman-Enskog technique to get the Navier-Stokes equations. In other words, the present UGKF naturally recovers the hydrodynamic flow behavior with the scale change.  

We now turn to the opposite case, $\delta_h\ll 1$, where the observation time scale is much shorter than the collision time. In this regime, collisions are rare and the flow lies in free-molecular regime. Particularly, it is easy to show that
\begin{equation}
	e^{-\nu_m h}=e^{-\delta_h} \approx 1-\delta_h, \quad e^{-\nu_m t} \approx 1-\delta_h (t/h).
\end{equation}
Then from Eq. \eqref{eq:f-beta} we can obtain
\begin{subequations}
	\begin{equation}
		f_\tF(\x,\bxi,t;h) = f(\x_0,\bxi,0) + O(\delta_h), 
	\end{equation}
	\begin{equation}
		f_\tT(\x,\bxi,t) = 0 +  O(\delta_h), \quad 
		f_\tC(\x,\bxi,t) = 0 +  O(\delta_h).
	\end{equation}
\end{subequations}
These results indicate that, in this limit, the sub-kinetic equation \eqref{eq:fF} for free-transport molecules dominates the system \eqref{eq:3Population}, and $f\approx f_{\tF}$, recovering the collisionless Boltzmann equation.

It is particularly interesting that $f_\tT(\x,\bxi,t) \to 0$ in both continuum and free-molecular limits, indicating that transitional molecules make negligible contributions in these regimes. In contrast, when $\delta_h \sim O(1)$, corresponding to the transitional regime, the contributions of all of the three populations are comparable.

\subsection{A reformulated UGKF}
The above analysis shows that the present UGKF \eqref{eq:3Population} constitutes a unified gas kinetic model capable of describing transport processes continuously from kinetic to hydrodynamic scales. Particularly,
since Eq. \eqref{eq:fC} governs molecules that have already undergone frequent collisions, it is natural to replace this detailed kinetic description with the corresponding hydrodynamic model for their collective behavior. Specifically, the moments of Eq. \eqref{eq:fC} can be approximated by a set of the Navier–Stokes-level equations without altering the multiscale character of the framework, 
\begin{equation}
	\label{eq:Wc}
	\left\{
	\begin{aligned}
		&\partial_t\W_\tC+\nabla \cdot \F_\tC =\bm{S}, \quad 0 < t \le h, \\
		& \W_{\tC}(\x,0) = 0,\\
	\end{aligned}
	\right.
\end{equation}
where $\W_\tC=\int{\bm{\psi} f_\tC d\bxi}$ represents the conservative hydrodynamic variables, $\F_\tC=\int{\bxi\bm{\psi} f_\tC^{ns} d\bxi}$ denotes the hydrodynamic flux obtained from an approximation of $f_\tC$ given by Eq. \eqref{eq:Fc-continuum} or \eqref{eq:Fc-Maxwell}, $\bm{S}=\int{\bm{\psi} \nu (f_\tF + f_\tT) d\bxi}$ is the source term accounting for newly generated collided molecules from the uncollided and transitional populations. This replacement circumvents the need to explicitly evaluate the complicated gain term  $Q^+$ while still capturing the correct macroscopic dynamics. 

With this modification, Eqs. \eqref{eq:fF}, \eqref{eq:fT}, and \eqref{eq:Wc} form a reduce UGKF that unifies kinetic and hydrodynamic descriptions within a single set of governing equations. For clarity, we summarize the reformulated UGKF as follows,
\begin{subequations}
	\label{eq:R3Population}
	\begin{equation}
		\label{eq:rfF}
		\left\{
		\begin{aligned}
			&\partial_t f_{\tF} +\bxi\cdot \nabla f_{\tF} = 0, \quad 0 < t \le h, \\
			& f_{\tF}(\x,\bxi,0;h) = \beta_h(\x,\bxi) f_0(\x,\bxi), \\
		\end{aligned}
		\right.
	\end{equation}
	\begin{equation}
		\label{eq:rfT}
		\left\{
		\begin{aligned}
			&\partial_t f_{\tT} +\bxi\cdot \nabla f_{\tT} = -\nu (f_{\tT}+f_{\tF}), \quad 0 < t \le h, \\
			& f_{\tT}(\x,\bxi,0;h) =\left[1-\beta_h(\x,\bxi)\right]f_0(\x,\bxi),\\
		\end{aligned}
		\right.
	\end{equation}
	\begin{equation}
		\label{eq:rWc}
		\left\{
		\begin{aligned}
			&\partial_t\W_\tC+\nabla \cdot \F_\tC =\int{\bm{\psi} \nu (f_\tF + f_\tT) d\bxi}, \quad 0 < t \le h, \\
			& \W_{\tC}(\x,0) = 0, \\
		\end{aligned}
		\right.
	\end{equation}
\end{subequations}
This reformulated model enables accurate prediction of flow phenomena across the full spectrum from continuum to rarefied regimes by consistently coupling microscopic particle dynamics with macroscopic transport. In this regard, Eq. \eqref{eq:R3Population} defines a set of gas-dynamic equations applicable from the kinetic (Boltzmann) to the hydrodynamic (Navier–Stokes) scales. It is emphasized that these equations are fundamentally distinct from classical extended hydrodynamics, such as the Burnett-type equations and Grad’s moment systems. The Burnett models are obtained by truncating the Chapman–Enskog expansion under the assumption of small Knudsen number (Kn), while the Grad’s equations are derived by truncating a Hermite (velocity-moment) expansion of the distribution function. In contrast, the reformulated UGKF invokes no Chapman–Enskog or moment truncation: it advances the population variables directly, preserves the exact conservation structure, is asymptotically consistent with Navier–Stokes equations in the hydrodynamic limit, and captures finite-Kn nonequilibrium effects without resorting to ad hoc closures or ill-posed higher-order gradients.

The reformulated UGKF \eqref{eq:R3Population} also provides a flexible foundation for constructing multiscale numerical schemes that preserve the correct asymptotic limits in both rarefied and continuum regimes. By modeling the strongly collided population hydrodynamically, the framework retains kinetic fidelity where necessary while achieving computational efficiency where continuum behavior prevails.

\subsection{A possible pathway toward Hilbert’s sixth problem}
By bridging kinetic and hydrodynamic descriptions within a single framework, the present UGKF (and reformulated UGKF) is not only well-suited for describing multiscale gas flows and designing practical multiscale numerical methods, but also opens new directions for theoretical exploration. In particular, the explicit separation of molecular populations and scale-dependent formulation provides a natural platform for examining fundamental questions at the interface of kinetic theory and continuum mechanics. 

In particular, the UGKF \eqref{eq:3Population} or the reformulated UGKF \eqref{eq:R3Population}, may offer fresh insight into Hilbert’s sixth problem \cite{ref:Hilbert1901,ref:Gorban2018}. In general, gas flows governed by the Boltzmann equation involves two sets of scales: the kinetic scales, characterized by the mean collision time $\tau_m$ and mean free path $\lambda$, and the hydrodynamic scales, defined by the wave time scale $t_0$ and characteristic flow length $L_0$. By introducing an intermediate observation time scale $h$ and its associated spatial scale $l \sim c_s h$ ($c_s$ is the sound speed), the flow behavior can be examined from an intermediate perspective, where the dominant physics depends on the dimensionless parameter $\delta_h$.

This scale-dependent UGKF enables the systematic transition between kinetic and hydrodynamic descriptions, offering a new route toward resolving Hilbert’s sixth problem, which calls for a rigorous mathematical connection between atomistic (kinetic) models and continuum mechanics. Beyond practical advantages for computation, the formulation carries significant theoretical value: by explicitly incorporating an intermediate observation scale and decomposing molecular populations accordingly, it provides a rigorous and transparent bridge between microscopic dynamics and macroscopic fluid behavior, advancing our understanding of the continuum limit from first principles.

\section{Numerical tests}
Based on the reformulated UGKF \eqref{eq:R3Population}, we have developed multiscale numerical methods that couple kinetic and hydrodynamic descriptions. A fully deterministic method  is detailed in Appendix \ref{App:numerical method}, Eqs. \eqref{eq:rfF} and \eqref{eq:rfT} are solved using the discrete velocity method, while Eq. \eqref{eq:rWc} is solved using the gas-kinetic scheme \cite{ref:XuBook}. For high-speed compressible flows, we further designed a hybrid wave–particle algorithm, in which the uncollided and transitional populations are treated by a stochastic particle method and the collided population by the gas-kinetic scheme.

A selection of benchmark tests and practical applications are presented to illustrate the effectiveness of the UGKF. The problems include the shock structure problem, the Couette flow in both continuum and transitional regimes, and external flows around the Mars Pathfinder forebody and X38-like re-entry vehicle. In these simulations, the numerical resolution itself provides the observation scale, with the time step $\Delta t$ and mesh size $\Delta x$ setting the temporal and spatial windows over which the flow physics is captured. 

In all simulations, a heuristic velocity-dependent collision frequency is adopted \cite{ref:beyond_Xu2021},
\begin{equation}
	\dfrac{\nu(\bxi)}{\nu_0}=
	\begin{cases}
		1, & \quad |C|\le 5\sqrt{RT},\\
		1 + 0.1 |C|/\sqrt{RT}, & \quad \mbox{otherwise},
	\end{cases}
\end{equation}
where $\nu_0=p/\mu$ with $\mu$ the dynamic viscosity. The dynamic viscosity can be evaluated as 
\begin{equation}
	\mu = \mu_\infty \left(\frac{T}{T_\infty}\right)^\omega,
\end{equation}
or given as a constant. Note that $\nu_0$ is also employed to replace $\nu_m$ in Eq. \eqref{eq:Fc-continuum} in the reconstruction $f_\tC$ from $\W$, either according to the Maxwellian or Shakhov distribution. 

The treatment of boundaries is similar to that in the discrete velocity method for the distribution function.
For instance, at a solid boundary the reflected part of the distribution function is a Maxwellian distribution that satisfies the non-penetration condition,
\begin{equation}
	\label{eq:BC}
	\int_{\bxi \cdot \bn >0} \bxi f_q (\x_w, \bxi, t) d\bxi + \int_{\bxi \cdot \bn < 0} \bxi f_M (\tilde{\rho}_q, \u_w, T_w) d\bxi = 0,
\end{equation}
where $\bn$ is the outward unit vector normal to the wall at point $\x_w$, $\u_w$ and $T_w$ are the wall velocity and temperature, respectively. From Eq. \eqref{eq:BC} the reflected density $\tilde{\rho}_q$ can then be determined.
This solid boundary treatment can be applied to all types of molecules (i.e., the subscript $q = F, T, C$).
Consequently, the boundary condition for the macroscopic flux of collisional molecules is given by
\begin{equation}
	\F_\tC(\x_w,t) =  \int_{\bxi\cdot\bm{n}>0}{\bxi\bm{\psi}(\bxi) f_\tC(\x_w,\bxi,t) d\bxi} + \int_{\bxi\cdot\bm{n}<0}{\bxi\bm{\psi}(\bxi) f_M(\tilde{\rho}_C, \u_w, T_w) d\bxi}
\end{equation}
For inlet or outlet (IO) boundaries, the ghost cell method is employed, and the entire distribution function in each ghost cell is provided at the beginning of each time step as an equilibrium state consistent with the boundary types.
Once these values are set in the ghost cells, the evolution near the IO boundaries is the same as that in the interior domains. 

Without further statement, the flow variables are normalized by the reference values, such as
\begin{equation}
	\hat{\rho} = \frac{\rho}{\rho_\infty}, \quad
	\hat{T} = \frac{T}{T_\infty}, \quad
	\hat{U} = \frac{U}{\sqrt{2RT_\infty}}
\end{equation}
The Knudsen number is defined as the ratio of mean free path over the reference length, ${\rm Kn} = \lambda_\infty / L_{ref}$,
where the mean free path $L_{ref}$ is computed by
\begin{equation}
	\lambda_\infty = \frac{16}{5} \sqrt{\frac{1}{2 \pi R T_\infty}} \frac{\mu_\infty}{\rho_\infty}
\end{equation}

\subsection{Shock structure at Mach 8}
\begin{figure}
	\centering
	\subfloat[]{\includegraphics[width=0.32\linewidth]{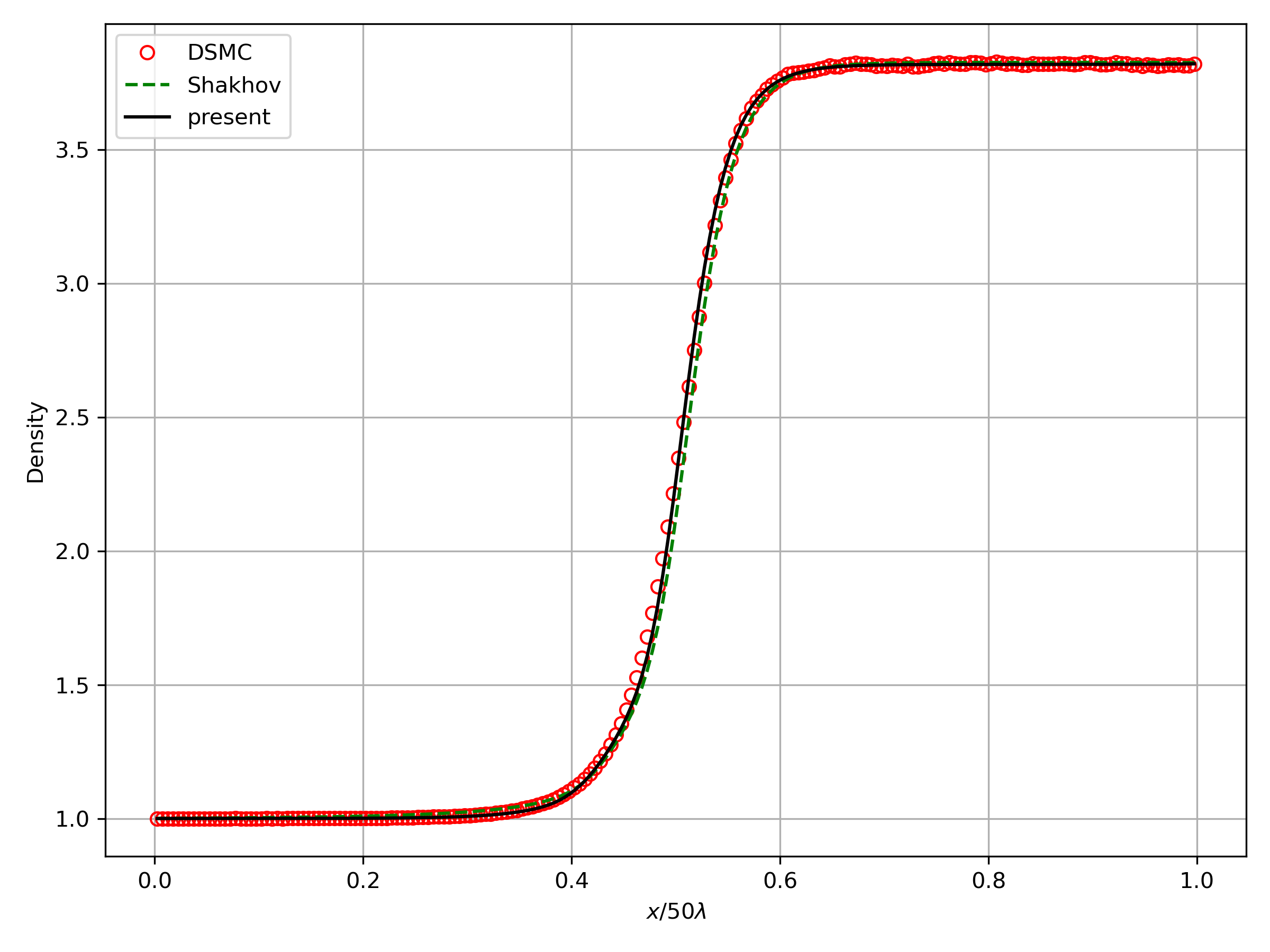}}
	\subfloat[]{\includegraphics[width=0.32\linewidth]{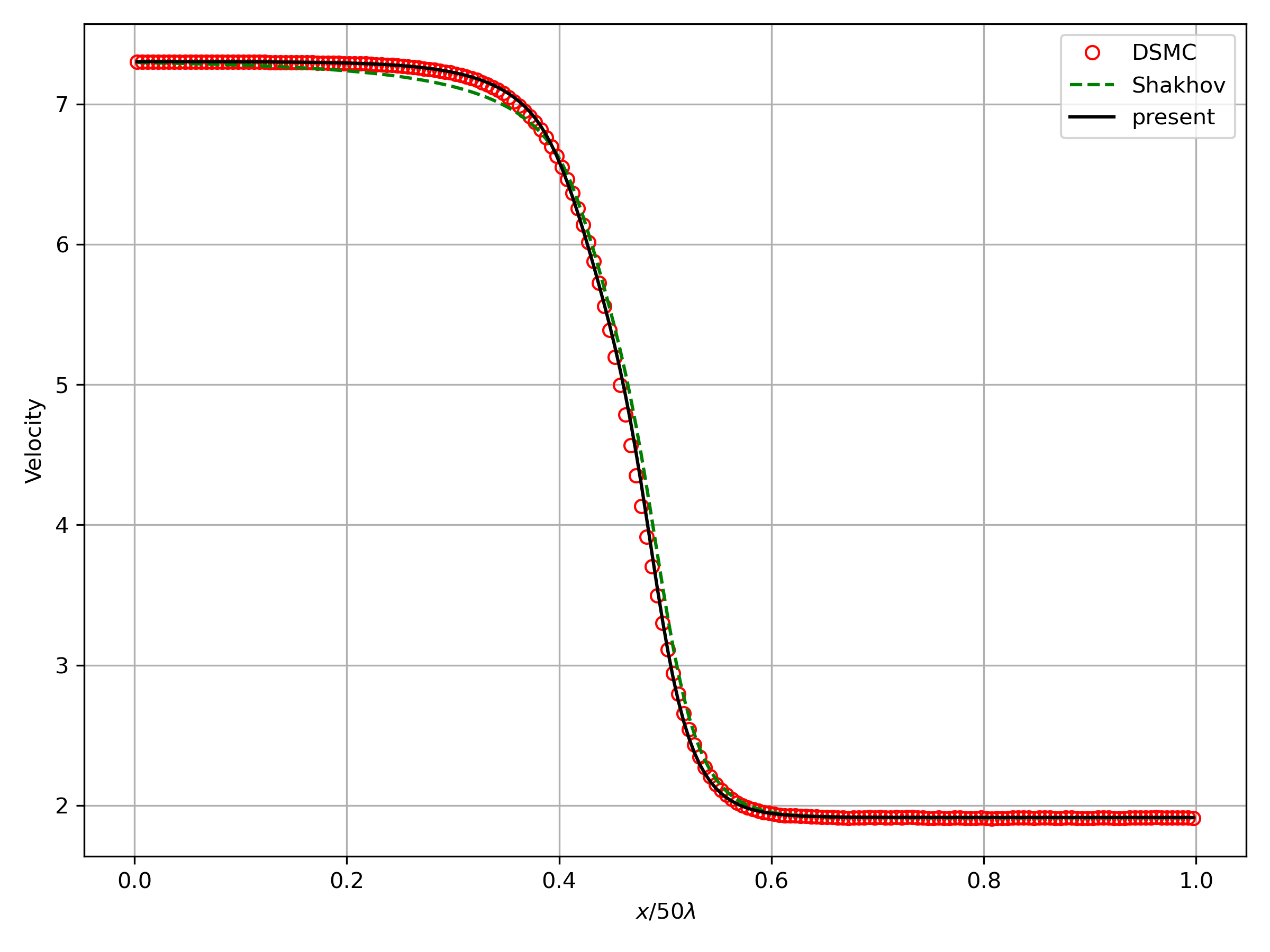}}
	\subfloat[]{\includegraphics[width=0.32\linewidth]{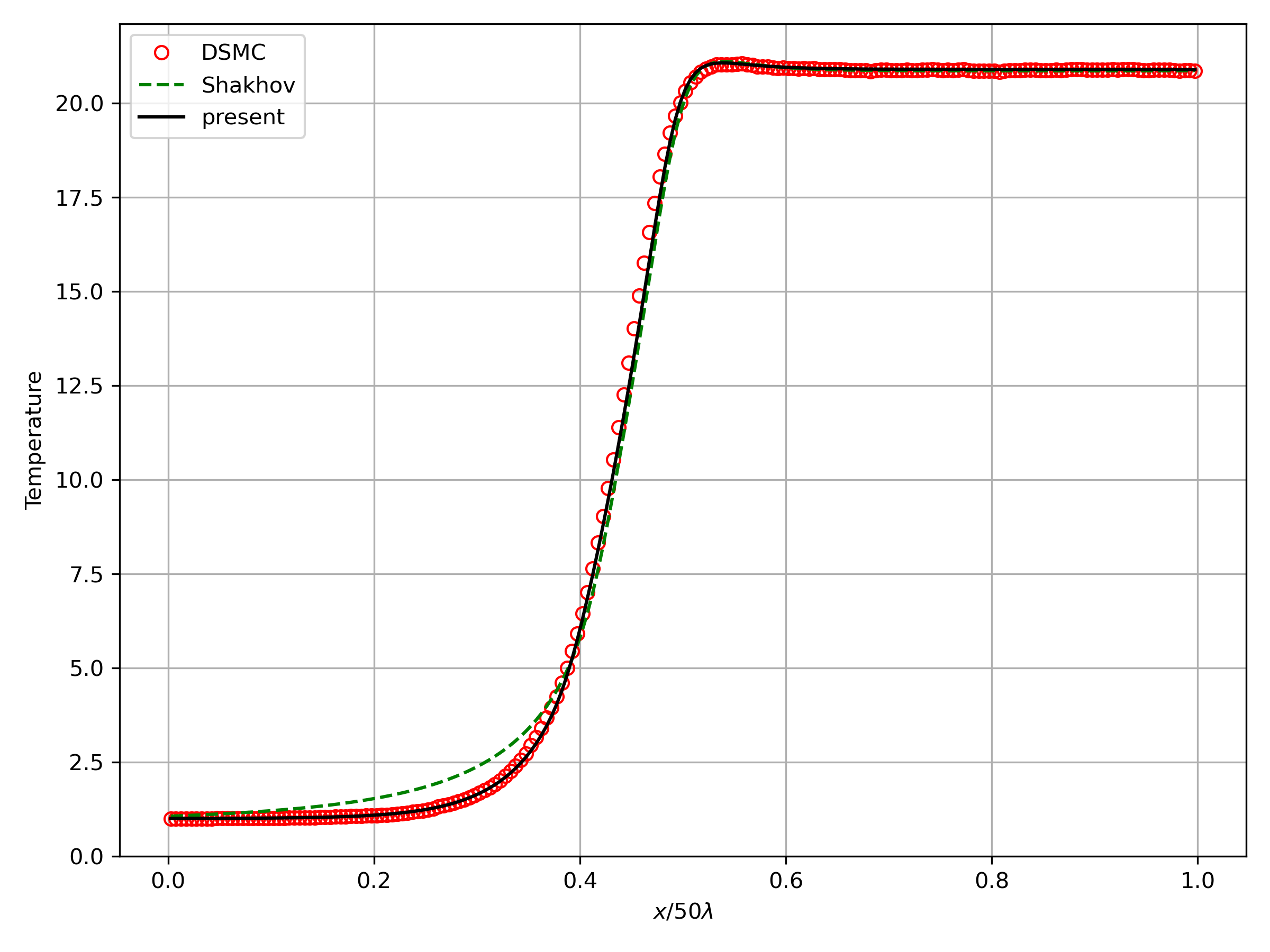}}
	\caption{Shock structure at Ma = 8: (a) density; (b) velocity; (c) temperature.}
	\label{fig:shock-structure}
\end{figure}

The first test is the shock structure at $\mbox{Ma}=8$ for a monatomic gas. 
The computational domain $[0,1]$ is discretized with a uniform mesh with 200 cells, and 
$600$ discrete velocity points spanning $[-15, 15]$ are adopted. The Knudsen number in the upstream is $0.02$, and the initial conditions in the left and right halves of the domain are given by
\begin{equation}
	(\rho, u, T)=
	\begin{cases}
		(1.0,7.303,1.0), & 0\le x<0.5,\\
		(3.8209, 1.9113, 20.872), & 0.5<x\le 1,
	\end{cases}
\end{equation}

The numerical results are shown in Fig.\ref{fig:shock-structure}, together with the DSMC data. It can be seen that the results of the present UGKF agree well with the reference data. Notably, the unphysical upstream temperature rises produced by classical BGK and Shakhov models \cite{ref:DUGKS15,ref:UGKS} is absent, demonstrating improved physical fidelity in capturing shock structures.
\subsection{Planar Couette flow}
We now consider the Couette flow in different regimes. The configuration consists of two vertical planar walls located at  $x=0$ and $x=L$, respectively. 

\begin{figure}
	\centering
	\subfloat[]{\includegraphics[width=0.45\textwidth]{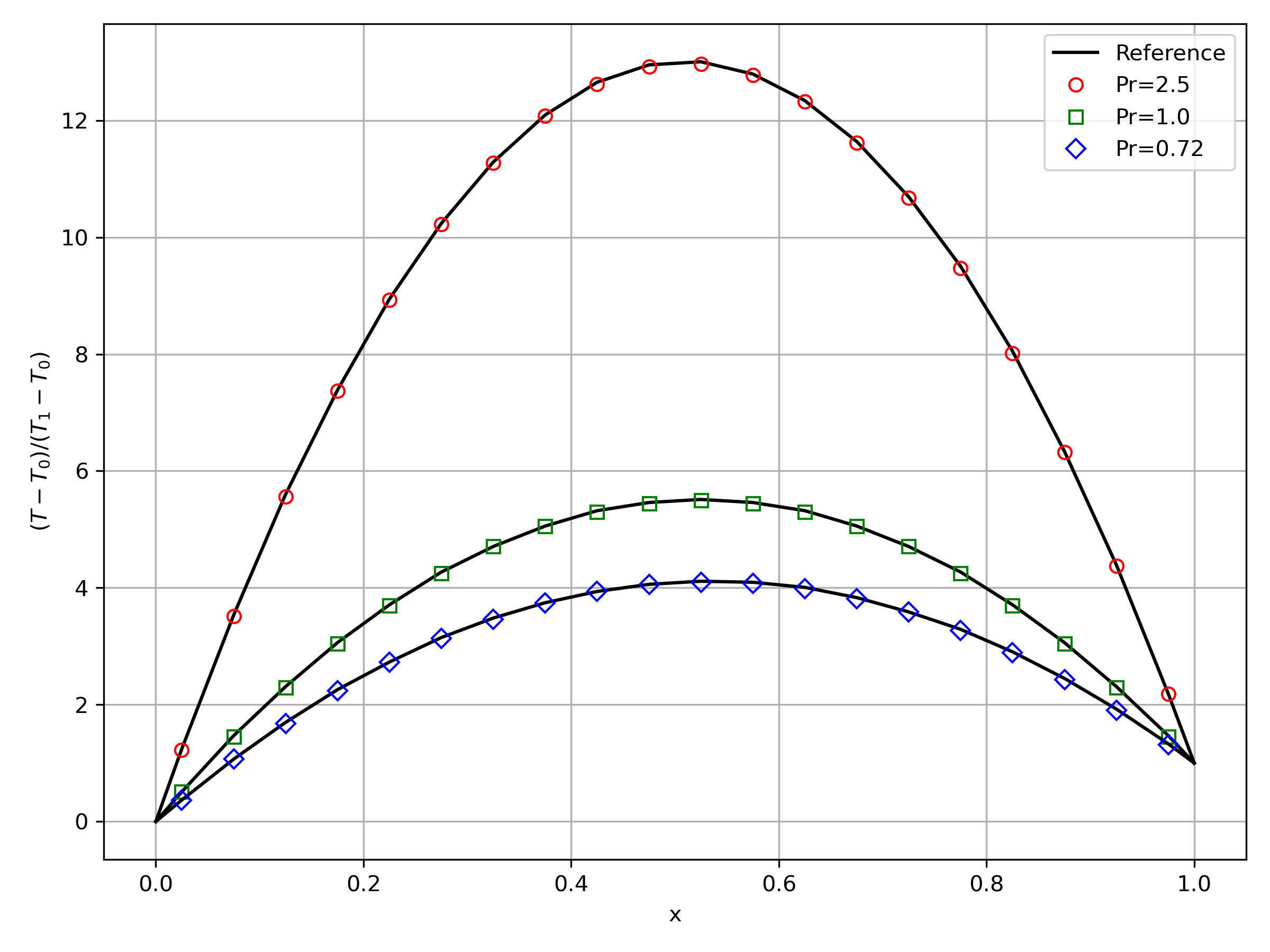}  
		\label{fig:CouetteA}}
	\subfloat[]{\includegraphics[width=0.45\textwidth]{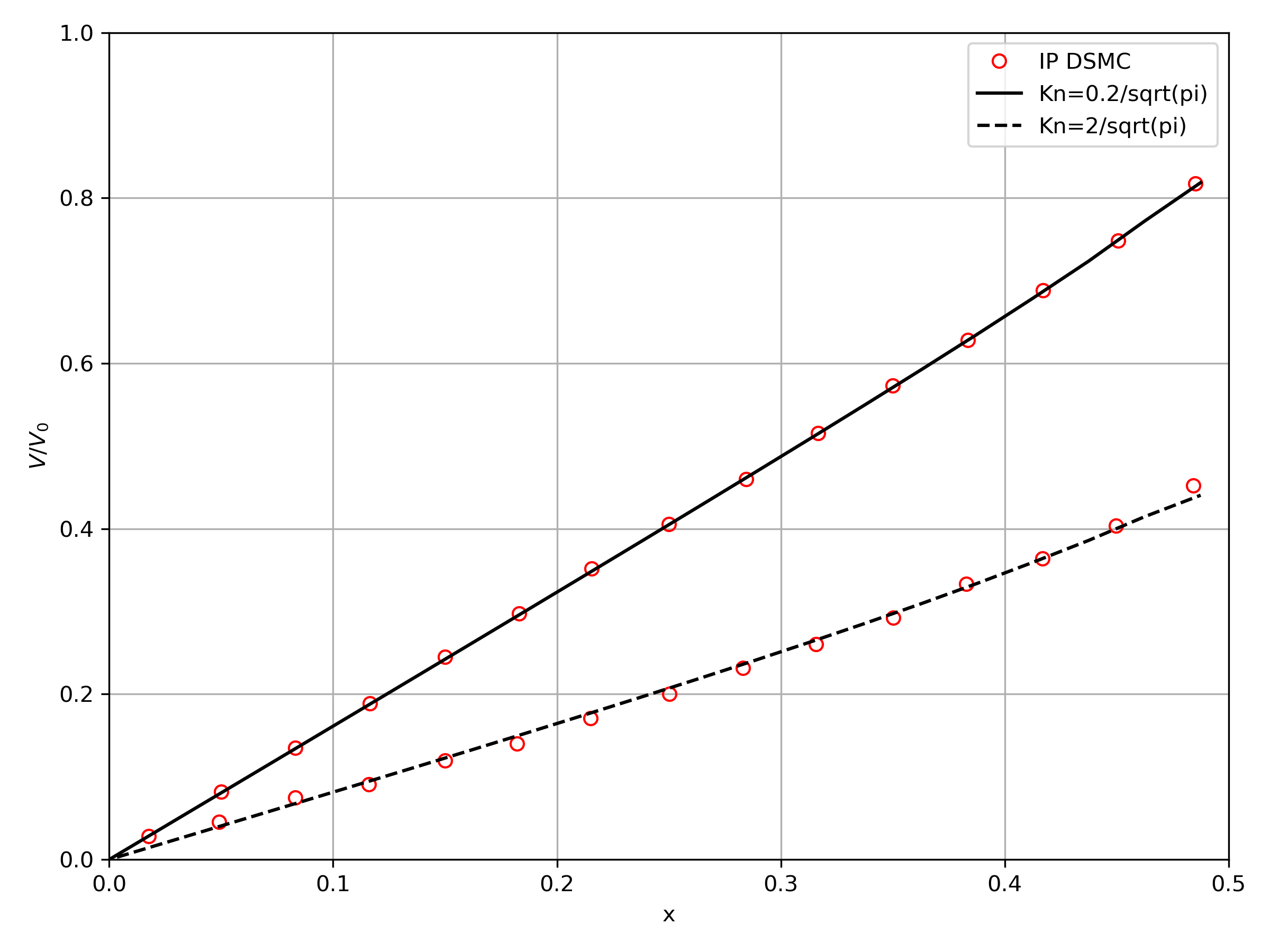}
		\label{fig:CouetteB}}
	\caption{continuum and transition Couette flow between two parallel plates. (a) Temperature at different Prandtl numbers, $\mbox{Kn}=10^{-5}$; (b) Velocity at different Knudsen numbers.}
	\label{fig:Couette}
\end{figure}

First, we simulate the flow in continuum limit, where the Knudsen number is chosen to be $\mbox{Kn}= 10^{-5}$. The left wall is stationary and maintains a fixed temperature $T_0$, while the right wall moves vertically with a constant speed $V_0$ and temperature $T_1$. 
With a constant dynamic viscosity, the temperature profile follows the following analytic solution,
\begin{equation}
	\dfrac{T -T_0}{T_1 - T_0} = x' + \dfrac{\rm Pr Ec}{2} x' (1-x'),
\end{equation}
where $x'=x/L$, $\rm Pr$ is the Prandtl number, $\rm Ec = V_0^2 / C_p(T_1 - T_0)$ is the Eckert number, with $C_p$ the heat capacity at constant pressure. The computational domain is discretized by a uniform mesh with 20 cells, and the velocity space is discretized into $80\times80$ points spanning $[-6,6]\times[-6,6]$. The moving velocity is set to $0.2$, corresponding to a low Mach incompressible flow. In this case, uncollided molecules contribute negligibly due to intensive collisions, and the moment equation for collided molecules dominates. The temperature profiles are shown in Fig.~\ref{fig:CouetteA}, which shows clearly that the present model successfully recovers the Navier-Stokes limit and yields accurate solutions at different Prandtl numbers.

We then consider the Couette flow in transitional regime. In this case, both walls are maintained at the same temperature and move oppositely with a constant speed $V_0 = 0.1$. The domain is discretized into a uniform mesh with $40$ cells, with the same velocity discretization as above. Two Knudsen numbers are considered, i.e., $\mbox{Kn}={0.2}/{\sqrt{\pi}}$ and ${2}/{\sqrt{\pi}}$. 
The velocity profiles predicted by the current model are shown in Fig.~\ref{fig:CouetteB}, together with DSMC data  \cite{ref:statistical_Fan2001}. It can be found that the non-equilibrium flow behavior in the transitional regime is also well captured by the current method, confirming the model's ability to capture non-equilibrium effects beyond the continuum description.

\subsection{Flow over the Mars Pathfinder forebody}
As a practical application, we simulate the flow around the forebody of the Mars Pathfinder probe using the hybrid wave-particle method. The geometry follows the experimental setup in ~\cite{bcexp2}, with flow conditions taken from the third case of Ref. \cite{bcdsmc}. In this configuration, the inflow is continuum while the wake becomes rarefied flow due to the sharp expansion at forebody shoulder. 

In the simulation, the free-stream conditions are $\rho_{\infty}=46.62\times 10^{-5}~\rm{kg/m^3}$, $U_{\infty}=1634~\mbox{m/s}$, $T_{\infty}=15.3~\mbox{K}$ and $\mbox{Ma}_{\infty}=20.5$. The wall temperature is fixed at $300~\mbox{K}$ following Ref.~\cite{bcdsmc}, and the angle of attack is varied as $0^{\circ}$, $10^{\circ}$, $20^{\circ}$ and $30^{\circ}$.
Nitrogen is used as the working gas. At $273~\mbox{K}$, the dynamic viscosity is $1.6579\times10^{-5}~\rm{Ns/m^2}$, and scales with temperature as $\mu\sim T^{\omega}$ with $\omega=0.74$. The Prandtl number is $\mbox{Pr}=0.72$. In the experiment, the viscosity coefficient is evaluated by combining Sutherland's law with a linear law. Under these conditions, the inflow Reynolds number is $36{,}265$, based on the base diameter of $50~\mathrm{mm}$.
The computational mesh contains 0.365 million cells, with the first layer height set to $3\times 10^{-4}~{\mathrm{m}}$. The simulation is advanced for 30,000 iterations, followed by an additional 10,000 steps for time averaging. A reference particle number of 100 per cell is employed, leading to a total of about 36.5 million particles.

\begin{figure}
	\centering
	\subfloat[]{\includegraphics[width=0.45\textwidth]{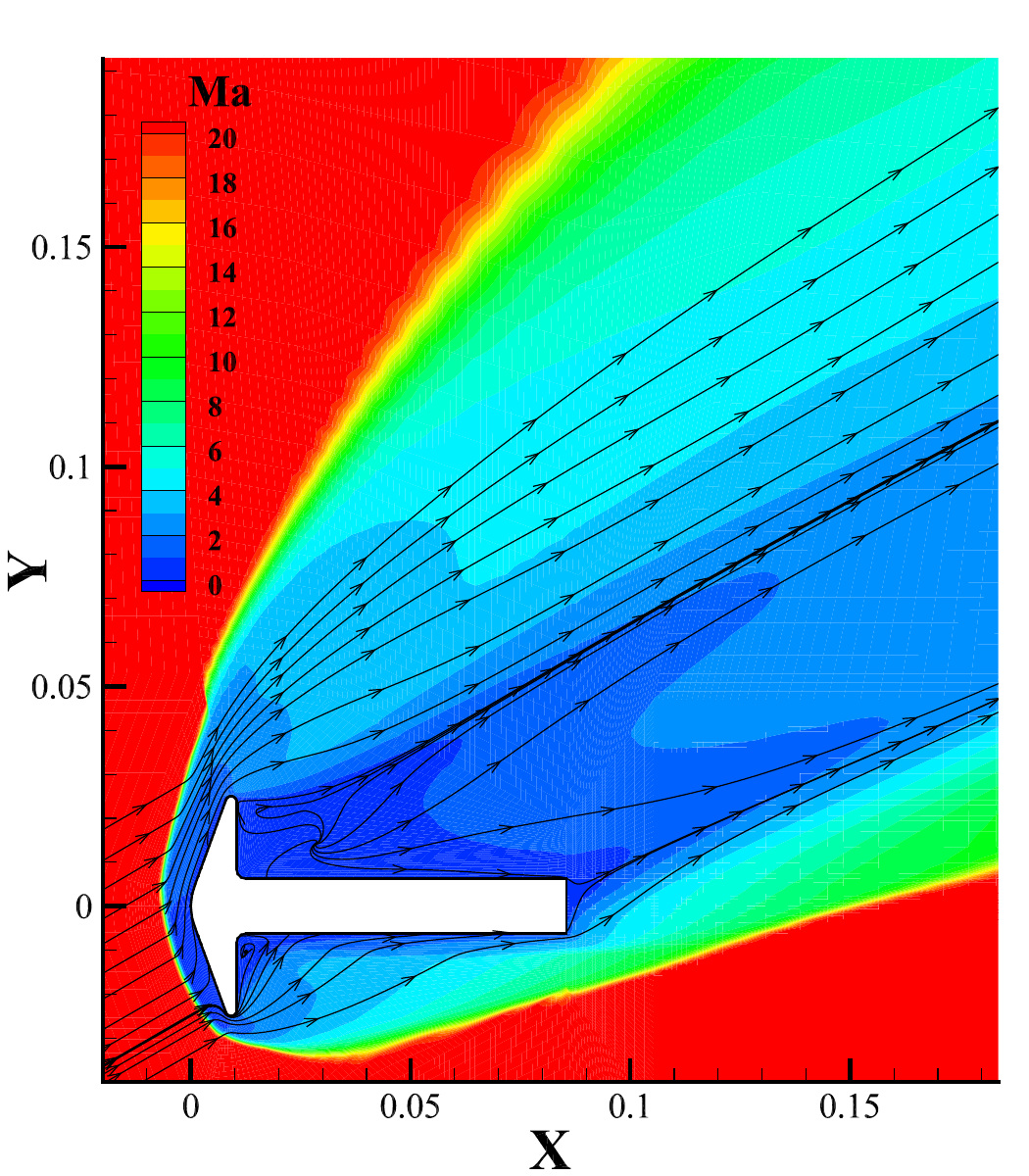}}
	\subfloat[]{\includegraphics[width=0.45\textwidth]{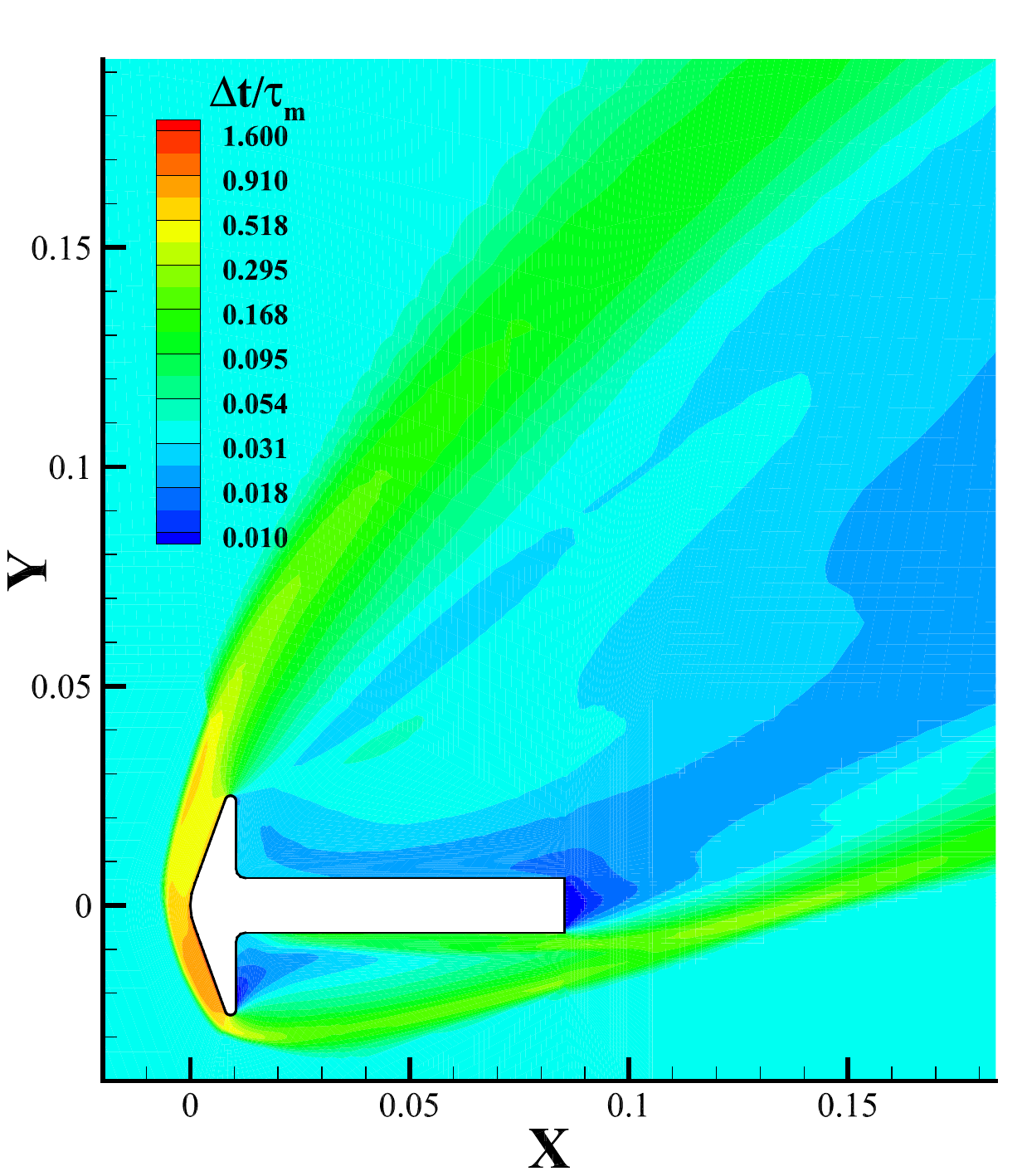}}
	\caption{Flow around the forebody of the Mars Pathfinde at $30^\circ$ angle attack.  (a) Local Mach number; (b) Local values of $\Delta t/\tau_m$.}
	\label{fig:Cone}
\end{figure}

\begin{figure}
	\centering
	\includegraphics[width=0.45\textwidth]{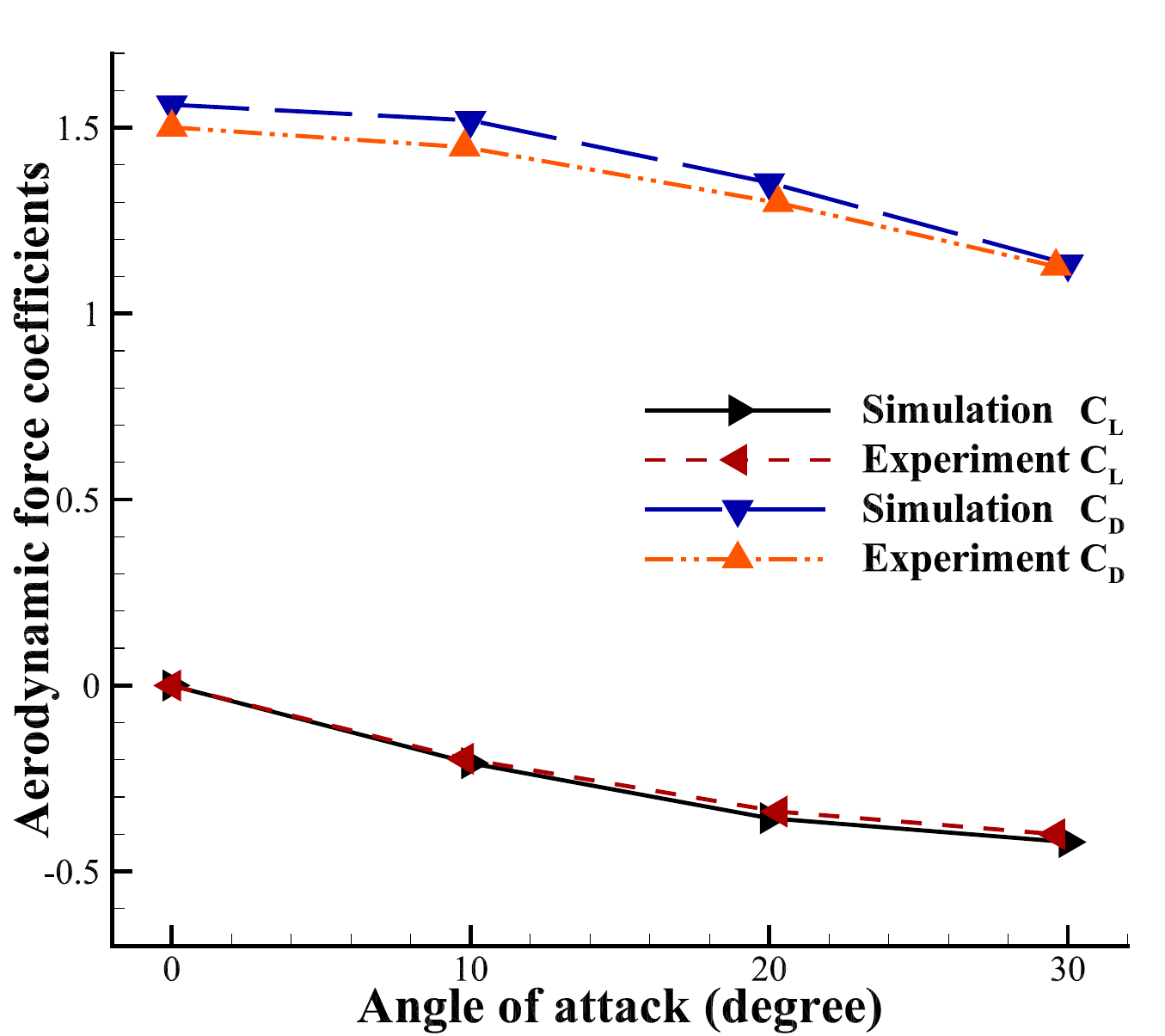}
	\caption{Drag ($C_d$) and lift ($C_L$) coefficients for the flow around the forebody of the Mars Pathfinde at angle attack $30^{\circ}$. Experimental data are taken from ~\cite{bcexp2}.}
	\label{fig:ConeForce} 
\end{figure}

Figure \ref{fig:Cone} shows the distributions of the local Mach number and $\Delta t/\tau_m$ around the forebody. The results demonstrate that the flow spans a wide speed range and exhibits clear multiscale features.
The predicted aerodynamic force coefficients (Fig.~\ref{fig:ConeForce}) agree quantitatively with experimental measurements \cite{bcexp2} over the full range of attack angles considered. By contrast, the Navier-Stokes solutions with slip boundary condition deviate significantly in the weak region from the DSMC results ~\cite{bcexp2,bcdsmc}, underscoring the improved physical fidelity of the present model.

\subsection{Flow around an X38-like vehicle}
As a second application, we simulate the flow around an X38-like vehicle \cite{ref:X38_DSMC}. 
The working fluid is Argon and the dynamic viscosity is modeled by the power law
\begin{equation}
	\mu = \mu_{ref}\left(\frac{T}{T_{ref}}\right)^\omega,
\end{equation}
where $\mu_{ref}$ is defined at the reference temperature $T_{ref} = 273\mbox{K}$. The flight altitude is about $78.6 \mathrm{km}$, with free-stream density $\rho_\infty = 1.11\times 10^{-4} ~\rm{kg/(m^3\cdot s)}$ and temperature $T_\infty = 56 ~\mbox{K}$. The free-stream Mach number and Knudsen number are $\mbox{Ma}_\infty=8.0$ and $\mbox{Kn}_\infty = 0.00275$, respectively. The angle of attack is set to $20^\circ$, and the wall temperature is $T_w=300 ~\mbox{K}$. The vehicle has a characteristic of $0.28 ~\mbox{m}$ and a reference area of $0.012 ~\mbox{m}^2$, respectively.
The aerodynamic coefficients, including pressure $C_p$, heat transfer $C_h$, and skin friction $C_\tau$, are normalized by the free-stream density $\rho_\infty$ and velocity $U_\infty$.

\begin{figure}
	\centering
	\subfloat[]{\includegraphics[width=0.45\textwidth]{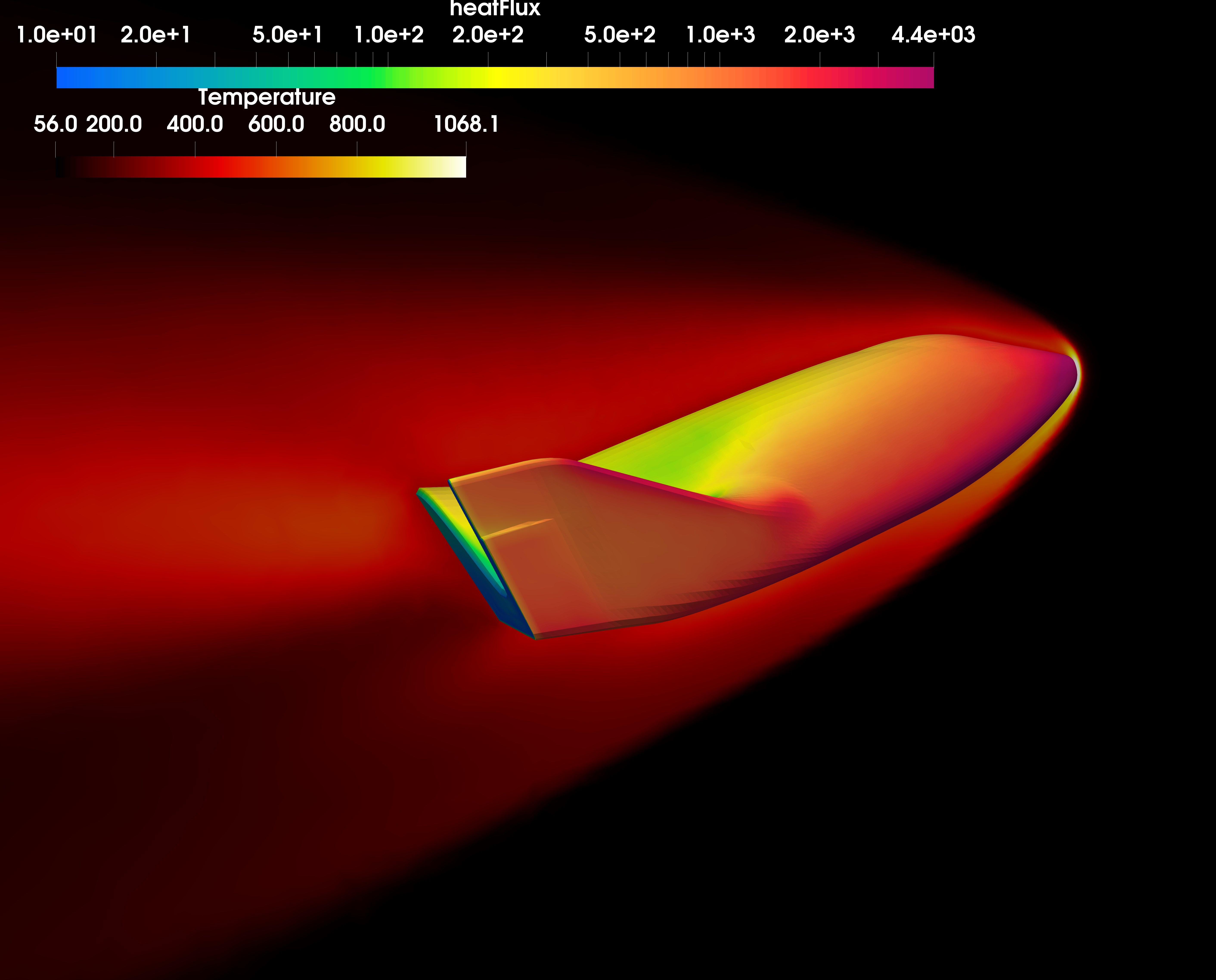}}
	\hfill
	\subfloat[]{\includegraphics[width=0.45\textwidth]{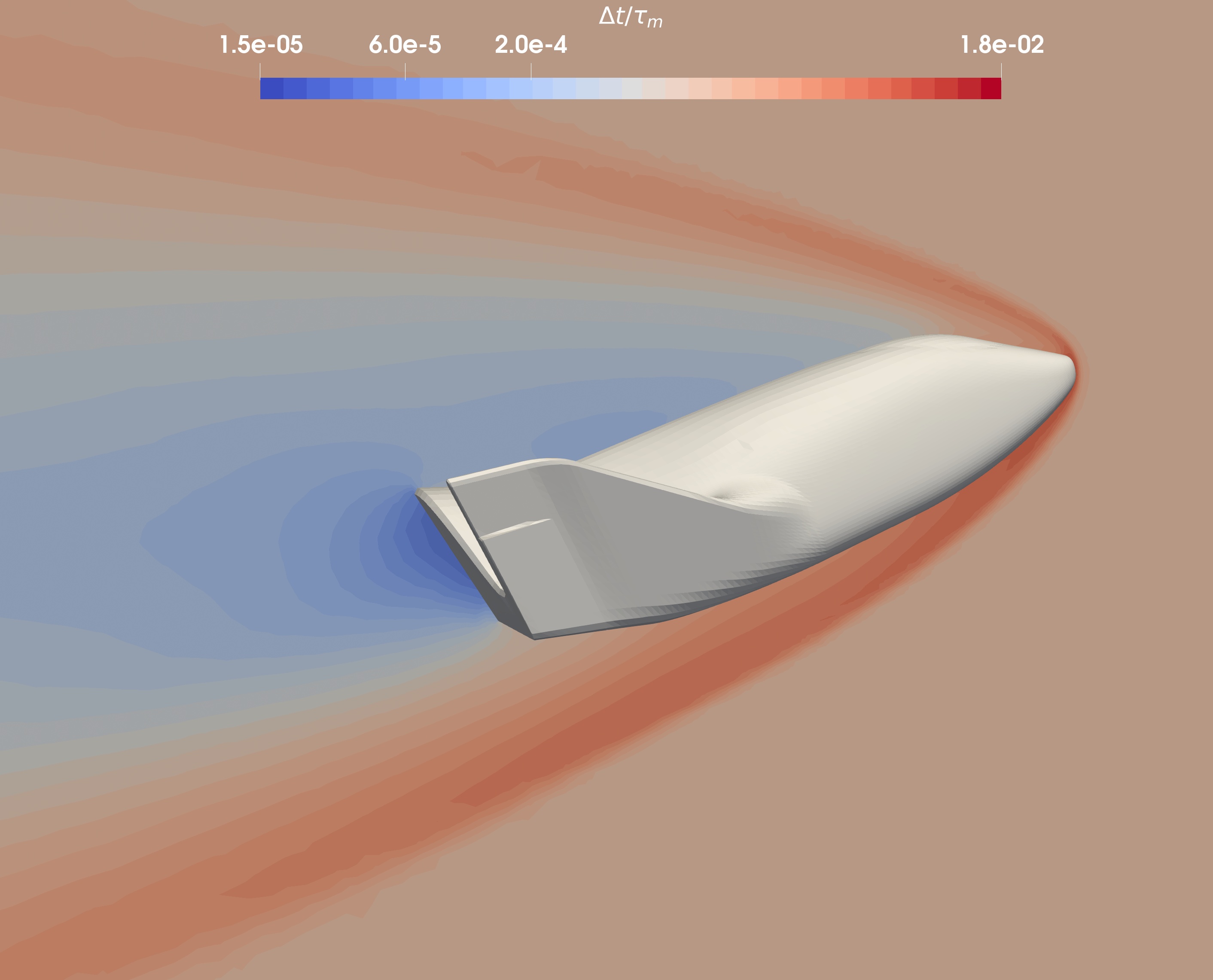}}
	\caption{Flow around the X38-like vehicle with $\mbox{Kn}=2.75\times 10^{-3}$ and $\mbox{Ma} = 8.0$ at angles of attack of $20^{\circ}$. (a) Heat flux and temperature,  (b) local values of $\Delta t/\tau_m$.}
	\label{fig:X38-1} 
\end{figure}

\begin{figure}
	\centering
	\subfloat[]{\includegraphics[width=0.32\linewidth]{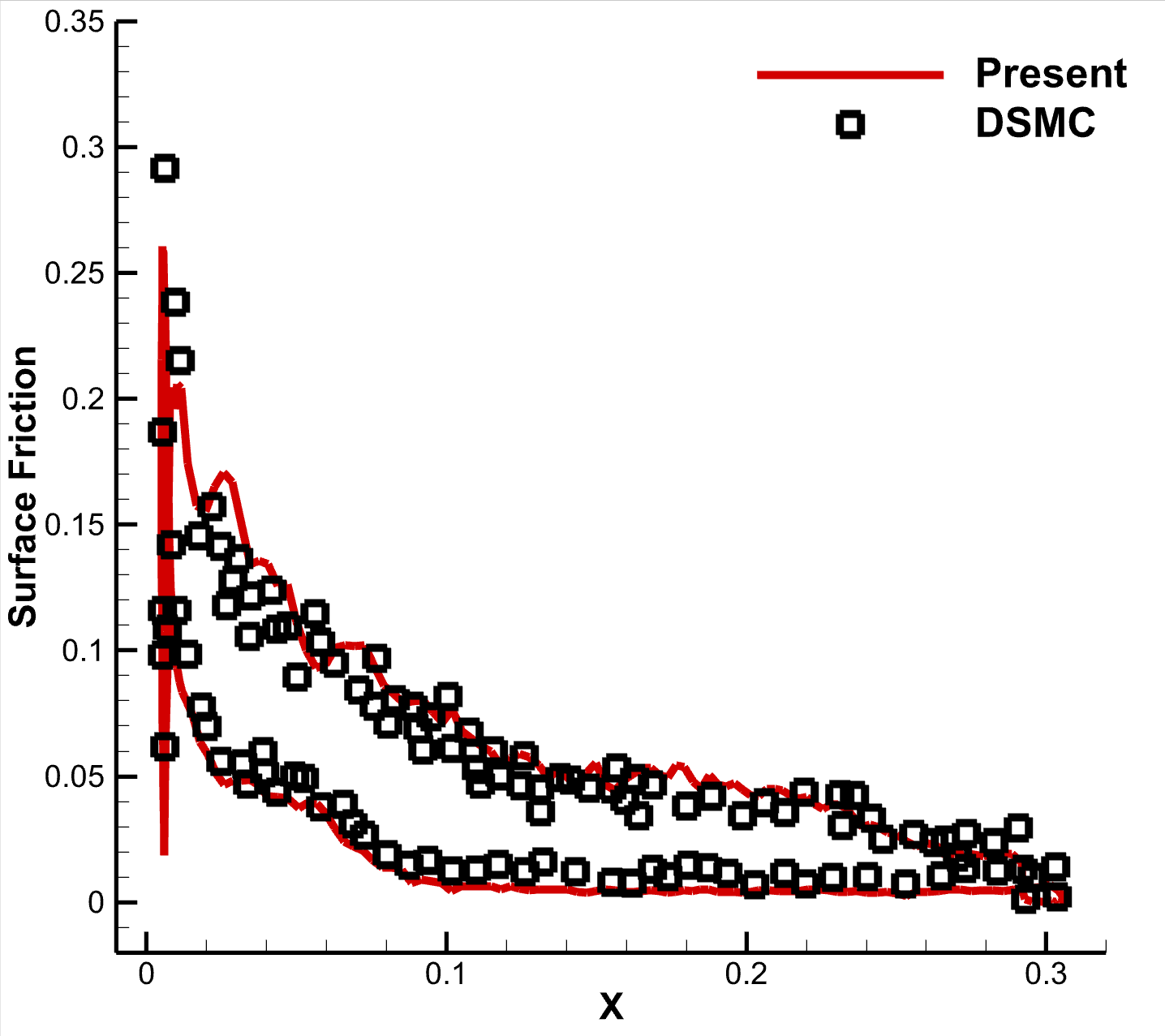}}
	\hfill
	\subfloat[]{\includegraphics[width=0.32\linewidth]{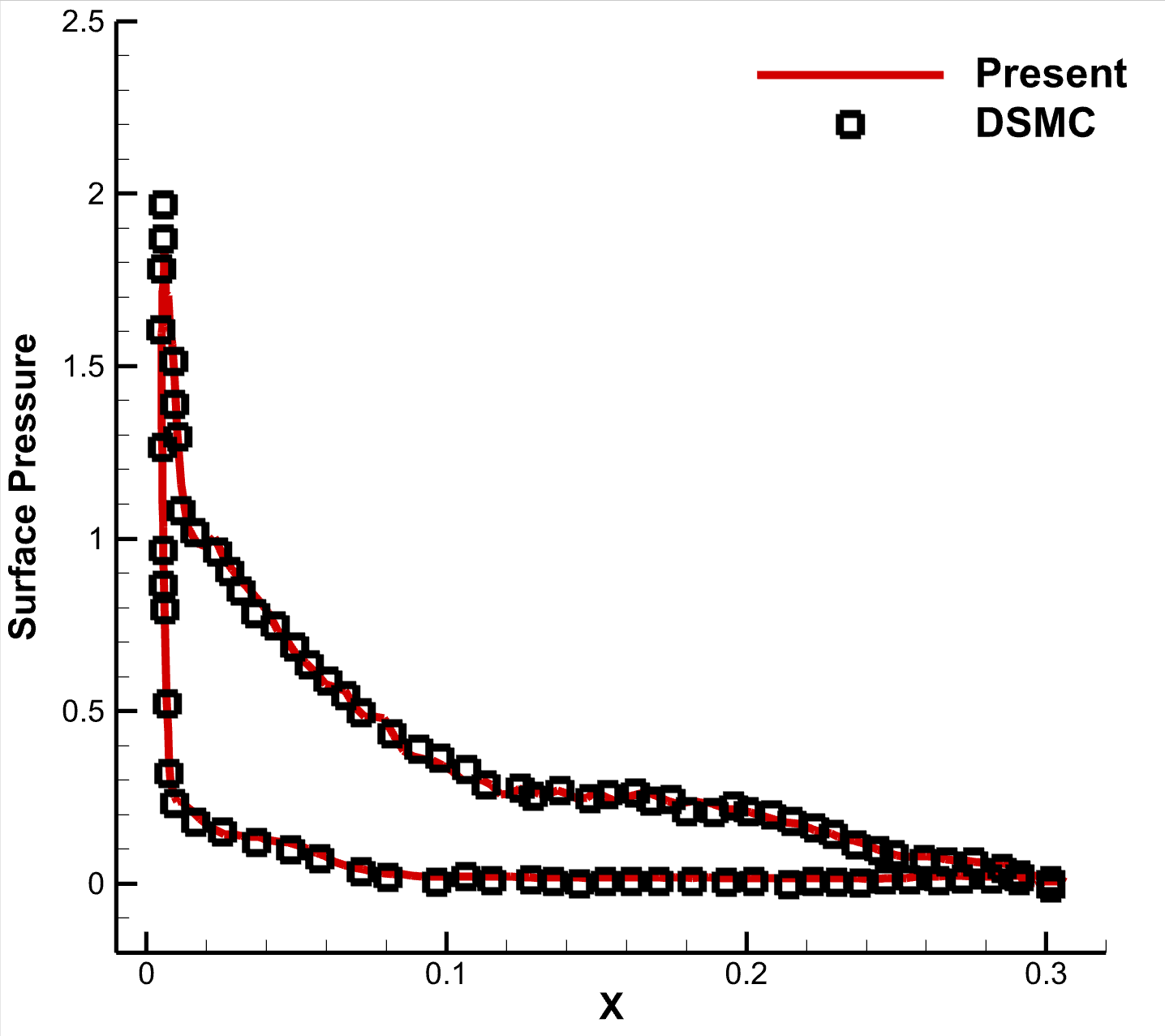}}
	\hfill
	\subfloat[]{\includegraphics[width=0.32\linewidth]{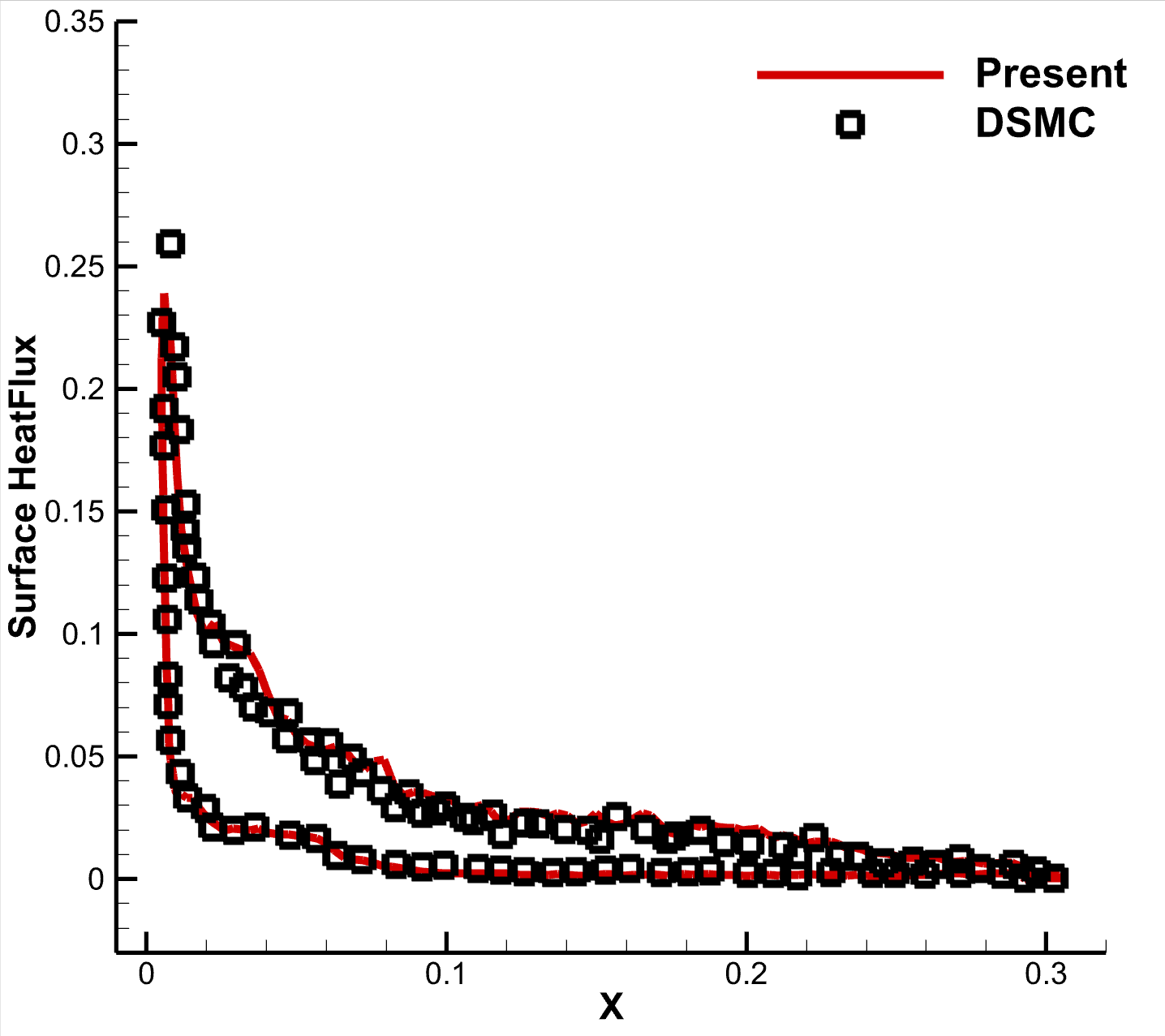}}
	\caption{Aerothermodynamic surface quantities of the flow around the X38 vehicle with $\mbox{Kn}=2.75\times 10^{-3}$. (a) skin friction, (b) surface pressure, and (c) surface heat flux. The DSMC data was taken from \cite{ref:X38_DSMC}.}
	\label{Sfig:X38-2} 
\end{figure}

Figure \ref{fig:X38-1} presents the distributions of surface heat flux, temperature, and the local value of $\Delta t/\tau_m$. The computed flow fields reveal multiscale structures across the entire domain, with local regimes ranging from continuum to rarefied. Although the free-steam Knudsen number is relatively low, consistent with a continuum regime, the value of $\Delta t/\tau_m$ varies by more than two orders of magnitude across the domain. This indicates that, at the mesh scales, the flow exhibits significant rarefaction effects in certain regions.
Figure~\ref{Sfig:X38-2} compares the computed aerodynamic forces and surface heat flux with DSMC results \cite{ref:X38_DSMC}, showing good agreement and confirming the predictive accuracy of the present model.

\section{Summary}
In this study, we developed a unified gas-kinetic framework for multiscale gas flows by classifying molecules, over a prescribed observation scale, into three populations according to their collision histories. From the formal integral solution of the Boltzmann equation, the sub-kinetic evolution equations (with corresponding initial data) for each population are derived. The resulting UGKF is mathematically equivalent to the Boltzmann equation yet exposes flow physics across variable scales beyond the mean free path and collision time.

Overall, the kinetic equations for molecules undergoing different collision behaviors present a new framework to describe gas evolution. Within this framework, the multiscale transport process can be revealed more directly and transparently. Moreover, the transition from the collisionless state to the collided state is clearly identified. Several non-equilibrium flows are simulated by a deterministic method and a hybrid continuum-particle method based on the kinetic framework, and the numerical results demonstrate that the present framework is indeed able to capture flow behaviors in different flow regimes.

In summary, the proposed UGKF provides a seamless connection between Boltzmann and Navier-Stokes solutions descriptions through a single set of governing equations. Based on the present framework, it may be more legitimate to answer Hilbert's 6th problem, where the different flow behavior must be associated with an observational scale from the kinetic to the hydrodynamic one through the variation of $h$. 
We remark that although the present kinetic framework is designed based on the Boltzmann equation for monatomic gases, it can also be extended straightforwardly to the multiscale transport of other particles, such as gas mixture, polyatomic gas, plasma, neutron, phonon, and photon. Particularly, it would be interesting to apply this framework to turbulent flows using specific kinetic descriptions, e.g. \cite{ref:Turb_Lundgren69,ref:Turb_Degon2002,ref:Turb_ChenHD2023,ref:Turb_WP}.

\begin{acknowledgments}
Z.L. Guo is supported by the National Natural Science Foundation of China (Grant No. 12472290) and the Interdiciplinary Research Program of HUST (2023JCYJ002), K. Xu is supported by Hong Kong research grant council (16208324,16301222), Y.J. Zhu is supported by Shanghai Magnolia Talent Plan Pujiang Project.
\end{acknowledgments}

\appendix
\section{Numerical method}
\label{App:numerical method}
A numerical method in finite-volume formulation is developed based on the kinetic framework \eqref{eq:3Population}, in which the sub-kinetic equations for the free transport and transitional gas molecules, Eqs.\eqref{eq:fF} and \eqref{eq:fT}, are solved by a discrete velocity method with a discrete velocity set $\{\bxi_k: k=1,2,\cdots, N\}$, while the moment equation for the collided molecules, Eq. \eqref{eq:Wc}, is solved using the Gas-Kinetic Scheme. 

For the collisionless and transitional gas molecules in Eq.\eqref{eq:fF} and Eq.\eqref{eq:fT},
they can be solved separately or in a united manner.
For simplicity, the uncollided gas molecules $f_\tU = f_\tF + f_\tT$ can be directly solved by the combined equation
\begin{equation}
	\left\{
	\begin{aligned}
		&\partial_t f_{\tU} +\bxi\cdot \nabla f_{\tU} =-\nu f_{\tU}, \quad 0 < t \le \Delta t, \\
		& f_{\tU}(\x,\bxi,0;h) =  f(\x,\bxi,0).
	\end{aligned}
	\right.
\end{equation}
which can be discretized by the integration over a cell $V_i$ centered at $\x_i$ from $t_n$ to $t_{n+1}$,
\begin{equation}
	\frac{f_{\tU,i,k}^{n+1} - f_{\tU,i,k}^n}{\Delta t} + \dfrac{1}{|V_i|}\sum_{j} (\bxi_{k} \cdot \bn_{ji}) \hat{f}_{\tU,ji,k} |S_{ji}| = -\nu f_{\tU,i,k}^{n+1}
\end{equation}
where $\Delta t=t_{n+1}-t_n$ is the time step, $f_{\tU,i,k}^n$ is the cell-averaged value of $f_\tU(\bxi_k,t_n)$, $\bn_{ji}$ is the outward unit vector normal to the interface $S_{ji}$ between cell $V_i$ and $V_j$, $|V_i|$ is the volume of cell $V_i$, and $|S_{ji}|$ is the area of the interface $S_{ji}$.
The flux across cell interface is evaluated using a second-order upwind scheme, 
\begin{equation}
	\hat{f}_{\tU,ji,k} = \frac{1}{\Delta t} \int_{0}^{\Delta t} f_{\tU,ji,k}(t) dt = \frac{1}{\Delta t} \int_{0}^{\Delta t} e^{-\nu t}\left(f_{\tU,ji,k}^n - \bxi_k \cdot \nabla f_{\tU,ji,k}^n t \right) dt,
\end{equation}
which is the time-averaged distribution function at cell interface. 
The interface value $f_{\tU,ji,k}^n$ and the spatial gradients $\nabla f_{\tU,ji,k}^n$ are obtained by spatial interpolation from the averaged values and gradients in the neighboring cells according to the upwind property,
$$
f_{\tU,ji,k}^n =
\begin{cases}
	f_{\tU,i,k} + ({\bm r}_{ji} - {\bm r}_i) \cdot \nabla f_{U,i,k}^n &\quad \text{if}\quad \bxi_k \cdot \bn_{ji} > 0\\
	f_{\tU,j,k} + ({\bm r}_{ji} - {\bm r}_j) \cdot \nabla f_{U,j,k}^n &\quad \text{if}\quad  \bxi_k \cdot \bn_{ji} < 0\\
\end{cases}
$$
Note that the integral solution to $D_t \phi = -\nu \phi$ along the characteristic line is used in the construction of the distribution function $\phi(t)$ on the interface, where $\phi$ can be $f_T$ or $f_U$ whose source term is non-zero.
This treatment is important especially for the cases using an explicit interface flux, which can incorporate the effect of source term into the flux evaluation.

For the collided gas molecules, the macroscopic flux $\F_c$ is modeled by the equilibrium part of the gas-kinetic scheme (GKS) \cite{ref:XuBook}, which gives the Navier-Stokes solutions in the continuum limit. Specifically, 
\begin{equation}
	\F_{\tC,ji} = \frac{1}{\Delta t} \int_0^{\Delta t} \int (\bxi \cdot \bn_{ji}) \left(c_0 g_{ji} + c_1 \bxi \cdot \nabla g_{ji} + c_2 \partial_t g_{ji}\right) {\bm \psi} d\bxi dt,
\end{equation}
where $g$ represents the target equilibrium distribution function, such as the Maxwellian or Shakhov distribution function, and the expressions of $c_0$, $c_1$, and $c_2$ can be found in Ref. \cite{ref:XuBook}. It is evident that the macroscopic flux $\F_\tC$ from $f_\tC$ is fully determined by the equilibrium state, which depends on the macroscopic variables and their spatial and temporal derivatives. Particularly, $\F_\tC$ recovers the Navier-Stokes flux in the continuum limit.

Once $\F_\tC$ is obtained, we can update the hydrodynamic variables $\W_\tC$,
\begin{equation}
	\frac{\W_{\tC,i}^{n+1} - \W_{\tC,i}^n}{\Delta t} + \dfrac{1}{|V_i|}\sum_{j} \F_{\tC,ji} |S_{ji}| = S_{i}^{n+1}.
\end{equation}
Alternatively, we can update the whole hydrodynamic variables $\W$, 
\begin{equation}
	\frac{\W_i^{n+1} - \W_i^n}{\Delta t} + \dfrac{1}{|V_i|}\sum_{j} (\F_{\tC,ji} + \F_{\tU,ji}) |S_{ji}| = 0,
\end{equation}
where 
\begin{equation}
	\F_{\tU,ji}=\sum_k (\bxi_k \cdot \bn_{ji}) w_k f_{\tU,ji}{\bm \psi(\bxi_k)}
\end{equation}
with $w_k$ being the quadrature weights associated with the discrete velocity $\bxi_k$. Up to this point, the distribution function $f_\tC$ for the collided molecules can be reconstructed according to Eq. \eqref{eq:Fc-continuum}. Particularly, the solutions of the BGK and Shakhov models can be recovered in the continuum limit, respectively, as $f_\tC$ is re-constructed from a local Maxwellian equilibrium or a Shakhov distribution, i.e., by setting $f^+=f_M$ or $f^+=f^S$ in Eq. \eqref{eq:Fc-continuum}.

\bibliography{KFrame}

\end{document}